\documentclass[twocolumn]{aastex6}
\usepackage{natbib}
\usepackage{amsmath,amsthm,amssymb,amsfonts}
\extrafloats{200}
 
\shorttitle{Formation of Clouds on Hot Jupiters}
\shortauthors{Powell et al.}

\begin{document}

\title{Formation of Silicate and Titanium Clouds on Hot Jupiters}

\author{Diana Powell\altaffilmark{1}, Xi Zhang\altaffilmark{2}, Peter Gao\altaffilmark{3,4}, and Vivien Parmentier \altaffilmark{5}}
\altaffiltext{1}{Department of Astronomy and Astrophysics, University of California, Santa Cruz, CA 95064; \href{mailto:dkpowell@ucsc.edu}{dkpowell@ucsc.edu} }
\altaffiltext{2}{Department of Earth and Planetary Sciences, University of California, Santa Cruz, CA 95064}
\altaffiltext{3}{Department of Astronomy, University of California, Berkeley, CA 94720}
\altaffiltext{4}{51 Pegasi b Fellow}
\altaffiltext{5}{Aix Marseille Univ, CNRS, LAM, Laboratoire d'Astrophysique de Marseille, Marseille, France}

\begin{abstract}
We present the first application of a bin-scheme microphysical and vertical transport model to determine the size distribution of titanium and silicate cloud particles in the atmospheres of hot Jupiters. We predict particle size distributions from first principles for a grid of planets at four representative equatorial longitudes, and investigate how observed cloud properties depend on the atmospheric thermal structure and vertical mixing.  The predicted size distributions are frequently bimodal and irregular in shape. There is a negative correlation between total cloud mass and equilibrium temperature as well as a positive correlation between total cloud mass and atmospheric mixing. The cloud properties on the east and west limbs show distinct differences that increase with increasing equilibrium temperature. Cloud opacities are roughly constant across a broad wavelength range with the exception of features in the mid-infrared. Forward scattering is found to be important across the same wavelength range. Using the fully resolved size distribution of cloud particles as opposed to a mean particle size has a distinct impact on the resultant cloud opacities. The particle size that contributes the most to the cloud opacity depends strongly on the cloud particle size distribution. We predict that it is unlikely that silicate or titanium clouds are responsible for the optical Rayleigh scattering slope seen in many hot Jupiters. We suggest that cloud opacities in emission may serve as sensitive tracers of the thermal state of a planet's deep interior through the existence or lack of a cold trap in the deep atmosphere. 
\end{abstract}


\section{Introduction}
Observations of exoplanet atmospheres have revealed damped spectral features in transmission---indicating the presence of an optically thick absorber of stellar photons \citep[e.g.,][]{2012MNRAS.422..753G, 2013MNRAS.436.2974G, 2013ApJ...774...95D, 2013ApJ...778..184J,2013ApJ...778..183L,2013ApJ...779..128M,2011MNRAS.416.1443S,2013MNRAS.436.2956S,2014ApJ...783....5S,2014ApJ...790..108F, 2016A&A...590A.100M,2016MNRAS.463..604M,2017MNRAS.470..742L}. This damping of spectral features has been attributed to the presence of clouds and hazes and is observed in a variety of exoplanets with well-characterized atmospheres \citep[e.g.,][]{2013A&A...559A..33C,2014Natur.505...69K,2014ApJ...794..155K,2014Natur.505...66K,2013ApJ...765..127F,2016Natur.529...59S,2016ApJ...823..109I}. Further studies of infrared phase curves reveal nightside emission that can be readily explained by the presence of clouds \citep[e.g.,][]{2016ApJ...823..122W, 2017AJ....153...68S}.

While clouds appear to be pervasive on exoplanets, the properties of these clouds can vary substantially for planets that are seemingly quite similar \citep[e.g.,][]{2016Natur.529...59S,2018AJ....155..150M}. An understanding of cloud properties, such as particle size distribution and composition is necessary to correctly interpret current and future observations. Hot Jupiters in particular have a comparative wealth of atmospheric data as they are relatively good targets for transmission spectroscopy. However, a thorough understanding of these planets requires a theoretical understanding of the clouds present in their atmospheres. Theoretical techniques will be particularly necessary in furthering our understanding of exoplanetary atmospheres with the advent of exquisite observational datasets from JWST \citep{2016SPIE.9904E..0EG}. It will be invaluable for observational programs to have a detailed theoretical framework able to give insight into an atmosphere's cloud properties before observation. The framework presented in this work is necessary for such theoretical insights. 
\\

\subsection{Previous Studies}

Previous studies have shown that condensational cloud and photochemical haze properties are strongly dependent on detailed planetary properties such as atmospheric irradiation, chemical composition, and dynamics \citep[e.g.,][]{2014Natur.505...69K}. The properties of clouds and hazes can further vary with composition and first order formation mechanisms, for instance, clouds that nucleate homogeneously, clouds that form efficiently only in the presence of seed particles, and hazes that form via photochemistry. Each of these factors influences the particle size distribution, which in turn has an influence on the inferences made from observations \citep[e.g.,][]{GRL:GRL19708,JGRD:JGRD2121,ZHANG199959}.

Solar system observations, especially in-situ measurements on Earth, have further shown that there are multiple modes in the cloud particle size distribution and that these modes vary throughout the atmosphere \citep[e.g.,][]{1994AtmRe..32..143K,2004GeoRL..3113108C}. Recently, simple bimodal particle size distributions have been proposed to interpret certain exoplanet observations as well \citep[e.g.,][]{2013MNRAS.432.2917P}. Multi-modal particle distributions tend to form due to differences in particle composition and formation process. Thus, while there are some indications of trends in cloud properties with equilibrium temperature/stellar irradiation \citep{2016ApJ...817L..16S,2016ApJ...826L..16H,2016ApJ...828...22P,2017ApJ...834...50B}, this remains a complex problem that requires a detailed understanding of cloud formation and related processes. 

There are several different forward and retrieval modeling techniques that are currently used to understand atmospheric properties despite the observational limitations imposed by the presence of clouds \citep{2013ApJ...775...33M,2015A&A...580A..12L,2016A&A...594A..48L,2012ApJ...754..135M,2013A&A...558A..91P,2008ApJ...675L.105H,2008A&A...485..547H,2017ApJ...847...32L}. Each of these previous works rely on one of three ways of understanding and parameterizing cloud properties: equilibrium cloud condensation modeling, grain chemistry (a subset of the larger field of cloud microphysics), or microphysical modeling of the coagulation of photochemical hazes.  

Equilibrium cloud condensation models use thermochemical equilibrium arguments to determine a planet's atmospheric composition and whether or not a certain species will energetically favor condensation and cloud formation. The vertical distribution of the resultant cloud particles can then be determined through a consideration of parameterized cloud particle sedimentation balanced by lofting due to vertical mixing \citep{ackerman-marley-2001}. This technique has been applied extensively to interpret observations of brown dwarfs and exoplanets \citep[e.g.,][]{saumon-etal-2012,2012ApJ...756..172M,2013ApJ...775...33M,2015ApJ...815..110M} and has been applied to hot Jupiters in 3D by \citet{2013A&A...558A..91P} to investigate the potential for a day-night cold trap to deplete TiO on the dayside of HD 209458b. A simplified version of this model was further used in \citet{2016ApJ...828...22P} to show that transitions in cloud composition as a function of effective temperature can explain the observed variations in Kepler exoplanet light curves. Simplified work in this vein has shown that the chosen size distribution has a distinct effect on the resulting spectra and that, for log-normal distributions, the largest particles in the distribution dominate the cloud's spectral contribution \citep{2015A&A...573A.122W}. By assuming that clouds are responsible for the Rayleigh scattering slope observed in the optical spectra of hot Jupiters, \citet{2015A&A...573A.122W} further predict the presence of a distinct silicate feature in the infrared that may be observable using JWST.

Grain chemistry microphysical cloud models treat cloud formation from a kinetics approach where both the growth and diminishment of cloud particles proceed via heterogeneous chemical reactions on the surface of grains. Recent work has additionally considered the impact of plasma physics on dust evolution in substellar atmospheres \citep{2017arXiv171207866S}. This framework was originally developed in great detail for brown dwarf atmospheres \citep{2001A&A...376..194H,2004A&A...423..657H,2008ApJ...675L.105H,2008A&A...485..547H,2003A&A...399..297W,2004A&A...414..335W,2006A&A...455..325H,2009AIPC.1094..572W,2011A&A...529A..44W} and has since been applied to hot Jupiter atmospheres and extended to 3D \citep{2015A&A...580A..12L,2016A&A...594A..48L,2016MNRAS.460..855H}. In this approach the cloud formation process is typically assumed to begin with the formation of TiO$_2$ seed particles in the upper atmosphere that settle downwards and act as sites of cloud formation for species such as MgSiO$_3$, Mg$_2$SiO$_4$, SiO$_2$, Al$_2$O$_3$, and Fe. These models have been used to study brown dwarf emission spectra \citep{2011A&A...529A..44W}, and have shown that a vertical gradient in cloud composition likely exists in brown dwarf atmospheres and in the atmospheres of comparable hot exoplanets. Recent work in 3D for hot Jupiters has further shown that two well studied and representative planets, HD 189733b and HD 209458b, could possess clouds in their atmospheres comprised of the same species thought to exist on brown dwarfs \citep{2015A&A...580A..12L,2016A&A...594A..48L}. This recent work also uncovers vertical and latitudinal variations in cloud composition due to atmospheric dynamics and global temperature differences. The model of HD 189733b was shown to have a deeper cloud deck in comparison to HD 209458b, consistent with the presence of more pronounced molecular features in its transmission spectra. 

An initial study of the coagulation of photochemical hazes in the upper atmospheres of hot Jupiters has shown that a consideration of these small lofted particles can reproduce the observed transmission spectra of HD 189733b \citep{2017ApJ...847...32L}. In this model, haze particles are injected into the top of the atmosphere and are allowed to coagulate. In particular, this work has successfully reproduced the Rayleigh slope at short wavelengths. 

Each of these three methods of modeling clouds in extrasolar atmospheres has advantages and disadvantages. Equilibrium cloud condensation models are not computationally intensive and can therefore be easily coupled with other atmospheric models. This technique, however, does not include the physical processes that govern cloud formation---namely the processes of nucleation, condensational growth, and evaporation, each with distinctive timescales and dependancies on planetary properties. The lack of detailed microphysics therefore limits the predictive power of this approach. Furthermore, these models require an assumed size distribution of cloud particles, which may skew inferences from observations. 

Grain chemistry models are highly detailed and have built-in chemistry calculations. However, these models can be difficult to generalize due to their reliance on specific nucleation pathways for cloud formation. These models adopt the moment method in numerics that requires a prescribed shape of the particle size distribution. In other words, these models are not able to predict the particle size distribution from first principles. Furthermore, this approach does not consider the influence of saturation vapor pressure over the particle surface due to particle curvature (the Kelvin effect) and particle mixture (the Raoult effect) \citep{seinfeld2006atmospheric}, both of which can alter the resultant cloud properties. 

Modeling of photochemical haze properties via coagulation can be used to determine the fully resolved haze particle size distribution. However, current work in this approach does not consider interaction with background gases via nucleation, condensational growth and evaporation. Once considered, these processes may have a substantial impact on the predicted size distributions. 

\subsection{A New Modeling Framework}

In order to resolve the cloud particle size distribution from first principles we need a model that relies on bin-scheme microphysics. In this work we present the first model of cloud formation on hot Jupiters from the perspective of bin-scheme cloud microphysics. This approach was pioneered on Earth where water clouds form primarily via heterogeneous nucleation and then evaporate or grow through condensation or coagulation \citep[e.g.,][]{nla.cat-vn946461}. The microphysical processes of nucleation, growth, evaporation, and coagulation have been applied to every planetary body in the solar system with a substantial atmosphere. In particular, bin-scheme microphysics has been used to reproduce and understand observations of sulfuric acid clouds on Venus \citep[e.g.,][]{gao2014bimodal}, CO$_2$ and water clouds on Mars \citep[e.g.,][]{1993Icar..102..261M,1999JGR...104.9043C}, hydrocarbon clouds and hazes on Titan \citep[e.g.,][]{2003Icar..162...94B,2004GeoRL..3117S07B,2006Icar..182..230B,2010Icar..210..832L,2011Icar..215..732L}, and hydrocarbon hazes on Pluto \citep[e.g.,][]{2017Icar..287..116G}. 

In the bin scheme approach, the particle size distribution is discretized into multiple bins according to size. Each bin of particles evolves freely and interacts with other bins. Therefore, there is no a-priori assumption of the particle size distribution. Bin-scheme microphysics is widely used in cloud formation models of Earth's atmosphere and is able to reproduce the multi-modal distributions of cloud particles. 

We use the one dimensional Community Aerosol and Radiation Model for Atmospheres \citep[CARMA;][]{turco1979one,1988JAtS...45.2123T} to conduct a detailed parameter space study of titanium and silicate clouds on hot Jupiters taking into account cloud microphysics. CARMA models the processes that govern cloud formation from first principles and therefore allows us to not only determine cloud properties for a wide range of parameters but also to test the assumptions used in other cloud modeling efforts. CARMA, like grain chemistry modeling, treats cloud formation as a kinetics process. Thus particle formation and growth in CARMA also depends on how long it takes for the condensate molecule, or some rate limiting precurser (e.g., SiO in MgSiO3), to diffuse to the particle. In this work we calculate cloud properties for four representative locations along the equator of hot Jupiters (the substellar point, east limb, antistellar point, and west limb) as these planets are three-dimensional with atmospheric thermal profiles that vary with location.

Our approach can be applied to the wealth of condensates that have been hypothesized to exist in hot Jupiter atmospheres by chemical equilibrium modeling \citep{burrows-sharp-1999,lodders-2002}. We choose MgSiO$_3$ and TiO$_2$ as our cloud species for this initial survey because silicate clouds are one of the more optically thick condensates \citep{2015A&A...573A.122W} and titanium is thought to often condense in hot Jupiter atmospheres with equilibrium temperatures less than $\sim$ 2000 K \citep{fortney-etal-2008,2016ApJ...828...22P,2017MNRAS.464.4247W} which is supported by a dearth of observed atmospheric TiO features \citep[e.g.,][]{2016Natur.529...59S}. Titanium clouds may also nucleate more easily than silicate clouds and could thus be a condensation nuclei for the growth of other cloud species. 

In Section \ref{intuit}, we give an overview of the theory used in our cloud model. In Section \ref{modeling}, we discuss our model and computational setup in detail. In Section \ref{timescales}, we introduce characteristic timescales of relevant processes in our model. In Section \ref{results}, we discuss the results of our model grid and place these results in context. In Section \ref{obs}, we discuss observational implications. We provide several conclusions and summarize our work in Section \ref{sum}.

\section{Theory}\label{intuit}
The universality of the microphysical processes handled by CARMA makes it a powerful tool that can simulate virtually any condensate in any atmosphere, provided certain physical properties are known. While the processes of microphysics are well studied, this work constitutes one of the first instances in which they have been applied to exoplanet atmospheres. We therefore provide a brief overview of the relevant processes and how they impact the formation of clouds in our model. For the specific equations that govern all of these processes in CARMA please see \citet{2018arXiv180206241G} Appendix A. 

\subsection{Overview of Cloud Microphysics}
Essential microphysical processes of cloud formation include nucleation, condensation, evaporation and coagulation. Nucleation refers to the initial phase change of a gaseous species to a solid or liquid state that starts the cloud formation process. Nucleation can occur either homogeneously or heterogeneously depending on the energy barrier associated with the process and the availability of seeds or cloud condensation nuclei (CCN). CCN may take many forms, such as meteorite dust, photochemical hazes, or other cloud species (e.g., Lee et al. 2018). The associated energy barrier depends on the atmospheric conditions as well as the specific properties of a species---in particular its surface tension and molecular weight. It is easier for species with low surface tension and molecular weight to form homogeneously than species with high surface tension and molecular weight.  Heterogeneous nucleation---the nucleation of one species onto a different species in either a solid or liquid state---tends to occur more efficiently than homogeneous nucleation when there are abundant seeds and if these seeds are favorable surfaces for the condensing species to nucleate on which further depends on the contact angle between the two species. In this work we treat the contact angle parameter as a nucleation efficiency parameter, similar to sticking efficiency in growth calculations, as it is otherwise not well known. In particular, we assume a low contact angle ($\sim 0.1 ^{\circ}$), therefore providing an upper limit on cloud formation.

Heterogeneous nucleation is the favored pathway for cloud formation in the case of water clouds on Earth \citep[e.g.,][]{nla.cat-vn946461}, CO$_2$ clouds on Mars \citep{1993Icar..102..261M,1999JGR...104.9043C}, ethane clouds on Titan \citep{2003Icar..162...94B,2004GeoRL..3117S07B,2006Icar..182..230B}, and sulfuric acid clouds on Venus \citep[e.g.,][]{gao2014bimodal}. Homogeneous nucleation, while less common in the solar system, is the favored pathway for the formation of high altitude water ice clouds on Earth \citep[e.g.,][]{GRL:GRL20901}. 

Once nucleation has occurred, the processes of condensational growth or evaporation can occur. Condensational growth allows a cloud particle to grow larger by many orders of magnitude. The pressure difference between the ambient gas pressure and the saturation pressure over the particle surface the driving force of both condensation and evaporation. Thus, many factors (such as temperature, curvature, and composition) could complicate the condensation and evaporation processes that fundamentally influence the final particle size distribution \citep[e.g.,][]{zhang2012}.

Cloud particles are also free to undergo coagulation, commonly modeled as Brownian coagulation on small scales and controlled by the random collisions among particles \citep[see][for the implementation used in CARMA]{2018arXiv180206241G}. We note that coagulation has been shown to play an important role in the evolution of photochemical hazes on Titan \citep[e.g.,][]{2010Icar..210..832L} and may be important in the evolution of high altitude photochemical hazes on hot Jupiters if haze is produced with an efficiency similar to that for Jupiter or Titan \citep{2017ApJ...847...32L}. However, given the relatively low number densities of large particles produced in our modeling, coagulation does not significantly change the resultant particle size distributions when fully included in our modeling procedure. The effect of coagulation has been tested in all simulations presented in this work. We therefore focus on the three dominant processes of nucleation, condensation, and evaporation throughout this work. 

\subsection{Governing Equations for Nucleation and Growth}\label{govern}
We apply classical theories of homogenous and heterogeneous nucleation to compute the rates of cloud particle generation \citep{nla.cat-vn946461,2011Icar..215..732L}. For homogenous nucleation the rate, in units of new particles per volume per unit time is, 
\begin{equation}
J_\text{hom} = 4 \pi a_c^2 \Phi Z n \exp(-F/kT),
\end{equation}
where $n$ is the number density of condensible vapor molecules, $k$ is the Boltzmann constant, and $T$ is temperature. The critical particle radius, $a_c$, is given by
\begin{equation}
a_c = \frac{2M\sigma_s}{\rho_p RT \ln S}
\end{equation}
where $M$, $\sigma_s$, $\rho_p$, and $S$ are the molecular weight, surface tension, mass density, and saturation ratio of the condensible species. $R$ is the universal gas constant. The energy of formation, $F$, is defined as

\begin{equation}
F = \frac{4}{3}\pi\sigma_s a_c^2.
\end{equation}

\noindent The rate of diffusion of vapor molecules to the forming particle, $\Phi$, in units of g cm$^{-2}$ s$^{-1}$ is given by

\begin{equation}
\Phi = \frac{p}{\sqrt{2\pi mkT}},
\end{equation}

\noindent where $p$ is the the partial pressure of the condensate vapor and $m$ is the mass of the vapor molecule. The inverse dependence on mass means that more massive molecules diffuse more slowly through the background gases. The Zeldovich factor, $Z$, takes into account non-equilibrium effects (such as the evaporation of newly formed particles) and is given by

\begin{equation}
Z = \sqrt{\frac{F}{3\pi kT g_m^2}},
\end{equation}

\noindent where $g_m$ is the number of molecules in particles of radius $a_c$.

The rate of heterogeneous nucleation, in units of critical germs per condensation nucleus, is given by,

\begin{equation}
J_\text{het} = 4 \pi^2 r_\text{CN}^2 a_c^2 \Phi c_\text{surf} Z \exp(-Ff/kT),
\end{equation}

\noindent where $r_\text{CN}$ is the radius of the condensation nuclei. The shape factor, $f$, is defined as

\begin{equation}
2f = 1+\left(\frac{1-\mu x}{\phi}\right)^3+x^3(2-3f_0+f_0^3)+3\mu x^2(f_0-1),
\end{equation}

\noindent where $\mu$ is the cosine of the contact angle between the condensible species and the nucleation surface, $x=r/a_c$, $\phi = \sqrt{1-2\mu x+x^2}$, and $f_0=(x-\mu)/\phi$. The number density of condensate molecules on the nucleating surface, $c_\text{surf}$, is given by

\begin{equation}
c_\text{surf} = \frac{\Phi}{\nu}\exp(F_\text{des}/kT),
\end{equation}

\noindent where $\nu$ is the oscillation frequency of the absorbed molecules on the nucleation surface, and $F_\text{des}$ is the desorption energy of that molecule. \citet{2018arXiv180206241G} gives a brief overview of typical $\nu$ and $F_\text{des}$ for different materials, however, the values for silicate clouds on titanium is not known. We therefore choose values typically chosen for water ($\nu = 10^{13}$ Hz, $F_\text{des} = 0.18$ eV), which \citet{2011Icar..215..732L} also used for hydrocarbons on tholin. To convert $J_\text{het}$ to units of newly nucleated particles per volume per time this quantity needs to be multiplied by the number of condensation nuclei.

The growth calculation in CARMA takes into account the diffusion of condensate particles to and away from the cloud particle, latent heat release, and several additional effects (see \citet{2018arXiv180206241G} and \citet{JACOBSON1994} for a full derivation of this process). The complete growth equation is defined as

\begin{equation}
\frac{dm_p}{dt} = \frac{4\pi r D' p_s(S-A_k)}{\frac{RT}{MF_v}+\frac{D'ML^2p_s}{k_a'RT^2F_t}},
\end{equation}

\noindent where $r$ is the size of the cloud particle, $p_s$ is the saturation vapor pressure of the condensate, $M$ is the condensate mean molecular weight, and $L$ is the latent heat of evaporation of the condensate. The ventilation factors, $F_v$ and $F_t$, account for the air density variations around a particle as it sediments in an atmosphere \citep{doi:10.1029/JD094iD09p11359,2011Icar..215..732L}. Note that the growth rate is directly proportional to particle size. 

The Kelvin factor, $A_k$, takes into account the curvature of a particle's surface and is given by

\begin{equation}
A_k = \exp\left(\frac{2M\sigma_s}{\rho_pRTr}\right).
\end{equation}

\noindent The molecular diffusion coefficient of the condensate vapor through the atmosphere, $D'$, and the thermal conductivity of the atmosphere, $k_a'$, are modified to account for gas kinetics near the particle surface and are defined as

\begin{equation}
D' = \frac{D}{1+\lambda Kn^c}
\end{equation}

\begin{equation}
k_a' = \frac{k_a}{1+\lambda_t Kn_t^c},
\end{equation}

\noindent where $\lambda$ and $\lambda_t$ are defined as

\begin{equation}
\lambda = \frac{1.33Kn^c + 0.71}{Kn^c+1}+\frac{4(1-\alpha_s)}{3\alpha_s}
\end{equation}

\begin{equation}
\lambda_t = \frac{1.33Kn_t^c + 0.71}{Kn_t^c+1}+\frac{4(1-\alpha_t)}{3\alpha_t}
\end{equation}

\noindent where $\alpha_s$ is the sticking coefficient and $\alpha_t$ is the thermal accommodation coefficient, which are both assumed to be order unity. The Knudsen numbers of the condensing gas with respect to the particle, $Kn^c$ and $Kn_t^c$, are given by

\begin{equation}
Kn^c = \frac{3D}{r}\sqrt{\frac{\pi M}{8 RT}}
\end{equation}

\begin{equation}
Kn_t^c = \frac{Kn^c k_a}{rD\rho_a(C_p-\frac{R}{2\mu _a})}
\end{equation}

\noindent where $C_p$ is the heat capacity of the particle, $\rho_a$ is the atmospheric mass density, and $\mu_a$ is the atmospheric mean molecular weight.

\subsection{Condensible Species}
For the purposes of this study, we consider the condensation of two species: MgSiO$_3$ and TiO$_2$. We note that many species are thought to condense at temperatures of $\sim$ 1000 - 2000 K. In particular, chemical equilibrium calculations show that other condensates such as Ti$_2$O$_3$, Ti$_3$O$_5$, MgAl$_2$O$_4$, Mg$_2$SiO$_4$, and CaTiO$_3$, among many others, may exist \citep{burrows-sharp-1999,lodders-2002}. We leave the investigation of other relevant cloud species to future work and instead focus on the wealth of information that can be understood more intuitively through the modeling of two species.

We choose MgSiO$_3$ because it is one of the most abundant cloud species in equilibrium cloud condensation modeling \citep{2017MNRAS.464.4247W}, evidence of silicate grain absorption has been observed on brown dwarfs \citep[e.g.,][]{cushing-etal-2006,2008ApJ...672.1159B,2008ApJ...686..528L}, and has a signature that could be seen with JWST/MIRI \citep{2015A&A...573A.122W}. MgSiO$_3$ has been proposed as a candidate for the Rayleigh scattering slope observed in transmission spectra due to its strong scattering properties \citep[see Section \ref{SSA};][]{2008A&A...481L..83L} though recent modeling of silicate clouds has called such assertions into question \citep{2017A&A...601A..22L}. We further use MgSiO$_3$ as a proxy for both Mg$_2$SiO$_4$ and MgSiO$_3$ as their optical properties are very similar, making them observationally difficult to distinguish \citep{2015A&A...573A.122W}, and because the reduced stoichiometry of MgSiO$_3$ makes its modeling more straightforward. 

We further consider the condensation of titanium in the form of TiO$_2$. We primarily consider TiO$_2$ due to its low surface tension, as explained in Section \ref{ST}. Titanium clouds are also appropriate candidate species because thermal inversions caused by TiO absorption \citep{burrows-etal-2007b,fortney-etal-2008} have not been observed in the majority of hot Jupiter atmospheres, suggesting that the titanium may have condensed out \citep{spiegel-etal-2009b,parmentier20133d,2016Natur.529...59S}. Indeed, TiO has only been observed for hot Jupiters with T$_\text{eq} >$ 2100 K \citep{2015ApJ...806..146H,2016ApJ...822L...4E,2017Natur.549..238S}, in line with theoretical predictions from \citet{fortney-etal-2008}. As such, we only consider cooler planets in this work.

\subsection{Assumptions Regarding Cloud Formation and Evolution}
Titanium and silicate clouds likely form via two different pathways. Titanium clouds are thought to commonly form via the following reaction:

\begin{equation}\tag{R1}
\text{TiO}_2 = \text{TiO}_2 \text{(s)},
\end{equation}

\noindent \citep{2006A&A...455..325H}. This reaction is a Type I reaction following the reasoning in \citet{2006A&A...455..325H} which is analogous to a gaseous molecule directly nucleating onto a grain. The seemingly direct nucleation and condensation of TiO$_2$ gas into solid TiO$_2$ cloud particles is well suited to modeling using classical nucleation and condensation theories without further assumptions. In our modeling we simply assume that all atmospheric Ti is located in condensible gaseous TiO$_2$.

Modeling the formation and evolution of silicate clouds requires additional assumptions due to uncertainties regarding their formation mechanism. We therefore adopt a simplified model for silicate clouds following classical formation theories. MgSiO$_3$ clouds are thought to form via the following reaction:

\begin{equation}\tag{R2}\label{sireac}
\text{Mg}+2\text{H}_2\text{O}+\text{SiO} = \text{MgSiO}_3 \text{(s,l)}+2\text{H}_2,
\end{equation}

\noindent \citep{2010ApJ...716.1060V}. In reality, it is likely that the three gases (Mg, H$_2$O, and SiO) will diffuse to the surface of a particle where they will undergo a reaction leading to nucleation or condensational growth. This is a Type III reaction in \citet{2006A&A...455..325H}, in which multiple gaseous species are involved. Following \citet{2006A&A...455..325H} Appendix B, we specify a key species (or educt) in the reaction, typically the least abundant species among the reactant molecules, that drives the surface reaction and growth \citep{2006A&A...455..325H}. For Equation \ref{sireac}, we choose SiO, as it is both the least abundant species assuming a solar composition gas \citep{lodders-2003} among the three molecules, and the heaviest, meaning that it takes the longest time to diffuse to the growing cloud particles.  We then assume that the cloud formation process is driven by the key species, SiO, such that MgSiO$_3$ cloud formation occurs when the partial pressure of SiO exceeds its equilibrium vapor pressure over MgSiO$_3$. Additionally, the formation of MgSiO$_3$ does not occur until an SiO molecule diffuses to the grain. 

Assuming a key species allows us to determine a reaction supersaturation ratio for silicate cloud formation that approximates formation via grain chemistry, defined as

\begin{equation}
S_r = S^{1/v_r^\text{key}},
\end{equation}

\noindent where $S_r$ is the reaction supersaturation ratio which gives the ratio of the growth and evaporation rates, $S$ is the standard supersaturation ratio, and $v_r^\text{key}$ is the stoichiometric factor of the key species in the reaction \citep{2006A&A...455..325H}. As our key species has a stoichiometric factor of unity, the reaction supersaturation ratio is the same as the standard supersaturation ratio. Finally, we assume that all atmospheric Si is present in the form of SiO. As this assumption tends to be roughly correct compared to actual elemental abundances to within an order of magnitude \citep{2010ApJ...716.1060V} we leave changes in abundance with temperature and additional cloud species to future work. Under these assumptions, classical nucleation and condensation theory can be used to approximate the microphysics of silicate cloud formation.

These assumptions and our general modeling scheme are not only similar to the scheme detailed in \citet{2006A&A...455..325H}, but are also analogous to earlier modeling of the formation of silicate dust in supernova remnants and stellar outflows \citep[e.g.,][]{2001MNRAS.325..726T}. More recent and detailed quantum chemistry calculations in the kinetic (as opposed to diffusive) regime have shown that actual nucleation rates may be suppressed at some temperatures and pressures and enhanced at high pressures compared to classical nucleation theory \citep{2018arXiv180304323M}. However, modeling at this level of detail is computationally expensive and outside of the scope of this work. We therefore adopt the above assumptions as a first step in understanding the formation of these complex clouds.

Finally, we do not consider radiative feedback of the clouds on the background atmospheric temperature structure and instead leave these calculations for future work.

\subsection{Surface Tension and the Kelvin Effect}\label{ST}
Our CARMA setup relies on the assumption that molecules react kinetically to form a species that can then nucleate or condense onto a cloud. This cloud microphysics approach, in which a condensible species forms and then nucleates or condenses onto a cloud, depends on the surface tension of each specific condensible species. 

In particular, the nucleation and condensation rates scale exponentially with surface tension to the third power (see Section \ref{govern}) such that species with larger surface tensions rarely nucleate homogeneously when CCN are present.

The surface tension of a species also governs its behavior with regard to heterogeneous nucleation and growth through the Kelvin effect, as described by the Kelvin equation

\begin{equation}
\ln\frac{p}{P_\text{sat}} = \frac{2\sigma V_\text{m}}{rRT},
\end{equation}

\noindent where $p$ is the vapor pressure over the particle surface, $P_\text{sat}$ is the saturation vapor pressure over a flat surface, $\sigma$ is the surface tension, $V_\text{m}$ is the molar volume, $r$ is the particle radius, $R$ is the universal gas constant, and $T$ is the temperature. 

Due to the Kelvin effect, the vapor pressure over the particle surface is larger than that on a flat surface and the effect depends on both surface tension and particle radius. For species with low surface tension the Kelvin effect is small, while for species with large surface tensions the Kelvin effect plays a role in the species's behavior with regards to growth and nucleation. The Kelvin effect causes species with large surface tensions to only heterogeneously nucleate or condense efficiently onto relatively large CCN or cloud particles with less curved surfaces. Furthermore, the Kelvin effect causes small particles to evaporate and large particles to grow with relative ease. 

TiO$_2$ has a surface tension of 480 erg cm$^{-2}$ \citep{2015AA...575A..11L} which is low enough for homogeneous nucleation to occur efficiently in our modeling. The low surface tension value also means that TiO$_2$ clouds are less susceptible to the Kelvin effect such that small cloud particles are less likely to evaporate once formed. TiO$_2$ can therefore produce both cloud particles and CCN that act as nucleation sites for other cloud species.  

The surface tension of magnesium silicate clouds is roughly 1280 erg cm$^{-2}$, measured in its solid state \citep[][for Mg$_2$SiO$_4$, where we assume the same value for MgSiO$_3$]{deLeeuw2000}. In our simulations of hot Jupiters we find that the supersaturation required for these clouds to homogeneously nucleate is extremely large. Therefore, if these clouds are abundant in hot Jupiter atmospheres, as suggested by equilibrium cloud condensation modeling, then their preferred method of formation must rely on heterogenous nucleation. We are thus forced to assume some form of CCN upon which heterogeneous nucleation can occur. For the purposes of this study, TiO$_2$ cloud particles act as the CCN. We note, however, that for MgSiO$_3$ cloud particles, growth is very efficient such that the Kelvin effect plays an insignificant role in determining the resultant cloud properties in our current modeling other than requiring silicate clouds to nucleate heterogeneously. This is because, regardless of the size of the initial CCN and evaporation of newly formed small cloud particles, silicate clouds will grow to roughly the same end size. 

\subsection{Transport Processes}\label{transpo}
Cloud particles are transported vertically in an atmosphere through the processes of gravitational settling and vertical mixing. Gravitational settling transports particles that form in the upper atmosphere to the lower atmosphere where they evaporate. Gravitational settling is modeled as Stokes fall velocity with a modifying Cunningham slip correction factor \citep[e.g.,][]{seinfeld2006atmospheric}. 

Turbulent vertical mixing in an atmosphere tends to decrease vertical gradients and smooth out inhomogeneities. Vertical mixing transports both gas and particles upward or downward depending on their relative mixing ratios. On hot Jupiters, the vertical mixing due to global circulation that consists of both upwellings and downwellings acts like a vertical diffusion process when globally averaged in a one-dimensional context \citep{2013A&A...558A..91P,2018arXiv180309149Z}. Vertical mixing in atmospheres is therefore often parameterized using a diffusion coefficient, K$_\text{zz}$, which encapsulates all vertical transport processes in an atmosphere such as vertical advection and vertical wave mixing. As recently demonstrated in \citet{2018arXiv180309149Z}, the global-mean eddy mixing on hot Jupiters should depend on the large-scale circulation strength, horizontal mixing and local cloud tracer sources and sinks due to microphysics. When K$_\text{zz}$ is large, an atmosphere is well mixed and diffusive transport is of increased importance.

A K$_\text{zz}$ profile cannot be directly derived from vertical velocities from 3D general circulation models without careful consideration of tracer transport as doing so results in an overestimated diffusivity \citep{2013A&A...558A..91P}. \citet{2018arXiv180309149Z} use a 3D GCM for hot Jupiters to show that different gaseous chemical species might have different eddy diffusion profiles, however, previous work in 3D from \citet{2013A&A...558A..91P} has demonstrated that the K$_\text{zz}$ parameter operates similarly for cloud particles of a broad range of sizes. 

\subsection{Atmospheric Cold Traps}\label{coldcold}
An atmospheric ``cold trap" can occur where the process of gravitational settling dominates the upward vertical mixing such that cloud particles rapidly settle after formation. In an atmosphere with a strong cold trap we expect to see the majority of cloud particles at the cloud base. This occurs because any cloud particles that form at higher altitudes will eventually settle downwards. At the same time, any gas that is vertically mixed upwards will first become supersaturated near the cloud base and will form clouds before reaching the upper atmosphere. 

If a species can become supersaturated at two points (i.e., the pressure and temperature profile crosses the condensation curve for a species at two points) in the atmosphere then it is possible for two cold traps to form. In this case, the lower cold trap is referred to as a ``deep cold trap". The deep cold trap may limit cloud formation in the upper atmosphere, therefore altering several atmospheric observables \citep[e.g.,][]{2013A&A...558A..91P,2016ApJ...828...22P}. Thus, the properties of clouds in the upper atmosphere can give insight into both the atmospheric vertical mixing and the deep thermal structure of a planet. 

In this paper we will determine the presence or lack of deep cold traps in an atmosphere as a way to understand how atmospheric observables may give insight into underlying planetary properties. 

\section{Modeling Approach}\label{modeling}
We adapt the Community Aerosol and Radiation Model for Atmospheres \citep[CARMA;][]{turco1979one,1988JAtS...45.2123T} version 3.0 \citep{JGRD:JGRD14488,JGRD:JGRD15781} for the study of titanium and silicate clouds on hot Jupiters.  We describe our model setup and adjustments to the base model in Section \ref{deets}. For a more comprehensive discussion of the microphysics and history of CARMA see \citet{2018arXiv180206241G} or \citet{turco1979one}, \citet{1988JAtS...45.2123T} and \citet{JACOBSON1994}. 

\subsection{Model Setup}\label{deets}
CARMA determines the quantitative effects of physical processes on cloud particle concentrations by solving a particle continuity equation. The processes included in our calculations are nucleation (both homogenous and heterogeneous), condensation and evaporation, sedimentation, and diffusion. The following continuity equation corresponds to these processes: 

\begin{equation}
\frac{\partial n}{\partial t} = \frac{\partial n}{\partial t}\bigg\rvert_\text{nuc.}+\frac{\partial n}{\partial t}\bigg\rvert_{\substack{\text{growth} \\ \text{ or evap.}}}+\frac{\partial n}{\partial t}\bigg\rvert_\text{sed.}+\frac{\partial n}{\partial t}\bigg\rvert_\text{diff.},
\end{equation}\label{continuity}

\noindent where $n$ is the cloud particle concentration, defined as $n(r,z,t)$ where $ndr$ is the number of cloud particles per volume of atmosphere at height $z$ with radii that range from $r$ to $r+dr$ at time $t$. The units of $n$ are particles cm$^{-3}$ $\mu$m$^{-1}$. A detailed discussion of each of these terms can be found in the appendix of \citep{2018arXiv180206241G} and is briefly discussed in Section \ref{intuit}. 

As noted before, CARMA operates using a bin scheme for particle microphysics where particle size is discretized into multiple bins that evolve freely and interact with other bins; this means that there is no a-priori assumption regarding the particle size distribution. 

We discuss our adaptation of CARMA to hot Jupiters in Sections \ref{cond} - \ref{vertmix}. A summary of the relevant model parameters can be found in Table \ref{mps}.

\begin{deluxetable*}{lll}
\tablecolumns{4}
\tablecaption{Model Parameters \label{mps}}
\tablehead{   
  \colhead{} &
  \colhead{Nominal Model} &
  \colhead{Other Values Used} 
}
\startdata
Surface Gravity     & 1000  cm s$^{-2}$& \\  
Atmospheric Mole. Wt. & 2.2 g mol$^{-1}$ (H/He) &  \\
Condensable Mole. Wt.   & 79.866 g mol$^{-1}$ (TiO$_2$) &  \\
  & 100.3887 g mol$^{-1}$ (MgSiO$_3$) &  \\
TiO$_2$ Surface Tension     &  480 erg cm$^{-2}$ \citep{2015AA...575A..11L} & \\
 MgSiO$_3$ Surface Tension    &  1280 erg cm$^{-2}$ \citep{deLeeuw2000} &  \\
T-P Profiles     &  Figure \ref{he_prof} (top panel) &  Figure \ref{he_prof} (bottom panel) \\
Diffusion Coefficient ($K_{zz}$) & 5$\times 10^8/\sqrt{P_{bar}}$  cm$^2$ s$^{-1}$ & 5$\times 10^7/\sqrt{P_{bar}}$,   5$\times 10^9/\sqrt{P_{bar}}$  cm$^2$ s$^{-1}$,  \\
& Constant at 5$\times 10^8$ below 1 bar & Constant at 5$\times 10^7$ and 5$\times 10^9$ below 1 bar \\
Time Step    &  100 s &  \\  
Total Simulation Time    &  $10^{9}$ s &  \\
Mass Ratio Between Bins  & 2 &     \\
Number of Bins    & 75 &     \\
Smallest Bin Size   & 1 nm &     \\
Largest Bin Size & 264 $\mu$m & \\
&&\\
   \textbf{Boundary Conditions}  & &     \\
Clouds (Top)  & Zero Flux &     \\
Condensation Nuclei (Top)  & Zero Flux &     \\
MgSiO$_3$ `Gas' (Top)  & Zero Flux &     \\
Clouds (Bottom)  & 0 cm$^{-3}$&     \\
Condensation Nuclei (Bottom)  & 0 cm$^{-3}$ &     \\
TiO$_2$ Gas (Bottom)  & Solar Abundance of Ti, $10^{-7.08}$ n$_\text{H}$  \citep{lodders-2003} &     \\
SiO Gas (Bottom)  & Solar Abundance of Si, $10^{-4.46}$ n$_\text{H}$  \citep{lodders-2003} &    \\
\enddata 
\end{deluxetable*}

\subsubsection{Saturation Vapor Pressures of Condensible Species}\label{cond}
For TiO$_2$, which exists in the gas phase, we use the saturation vapor pressure formula from \citet{2004A&A...414..335W}. In Equation \ref{tiosvp} we rewrite this formula in approximate form with pressure units of bar and temperature in Kelvin assuming solar metallicity of Ti which is contained in gaseous TiO$_2$. 

\begin{equation}\label{tiosvp}
P_\text{sat}  = 10^{(9.5489-(32450.8451/T )) }
\end{equation}

The condensation curves for each species are shown in comparison to the planetary pressure and temperature profiles in Figure \ref{he_prof} for the high and low entropy cases (described below). 

For MgSiO$_3$ we derive a condensation curve from \citet{2010ApJ...716.1060V} to calculate the saturation vapor pressure assuming that the limiting species for cloud formation is SiO. Condensation will occur when the partial pressure of SiO exceeds its equilibrium vapor pressure over MgSiO$_3$ \citep[cf. Table 3 in][]{2010ApJ...716.1060V}. We assume that all of the silicate in the atmosphere is locked up in SiO and use this to derive a partial pressure. The calculated condensation curve is given in Equation \ref{mgcond}, where [Fe/H] is the metallicity (which we take to be solar), and the saturation vapor pressure is given in Equation \ref{mgsvp}. In both equations pressure is in units of bar and temperature is in Kelvin. 

\begin{equation}\label{mgcond}
T(P_\text{total}) = \frac{10^4}{6.24-0.35\log_{10}(P_\text{total} )-0.7[\text{Fe/H}]}
\end{equation}

\begin{equation}\label{mgsvp}
P_\text{sat}  = 10^{(13.37 - 28571.43/T-[\text{Fe/H}] )}
\end{equation} 

\noindent This formulation assumes that only MgSiO$_3$ clouds form and neglects the formation of Mg$_2$SiO$_4$.

In our modeling we neglect changes in equilibrium elemental abundance with equilibrium temperature and instead assume a solar abundance in all cases. We can therefore expect the resultant cloud populations to represent an upper limit in mass.

\subsection{Planet Parameters and Grid}\label{planet}
We adapted CARMA for hot Jupiters through an adjustment of the surface gravity, atmospheric composition, the parameterized vertical mixing, and the pressure and temperature profile.

 \subsubsection{Pressure and Temperature Profiles}
 We use solar composition pressure and temperature profiles from \citet{2016ApJ...828...22P} without TiO/VO absorption for a Jupiter-size planet tidally locked around a solar-type star with gravity of 10 m s$^{-2}$ calculated using the SPARC/MITgcm \citep{showman-etal-2009}, a 3D general circulation model that uses the plane-parallel radiative transfer code of \citet{1999Icar..138..268M}. We run a grid of models with different equilibrium temperatures (T$_{\rm eq}$). Each planet in the grid has a unique T$_{\rm eq}$, semi-major axis, and planetary rotation rate accordingly. We consider 9 different T$_{\rm eq}$ (1300, 1400, 1500, 1600, 1700, 1800, 1900, 2000, and 2100 K) at 4 characteristic points in the atmosphere of a hot Jupiter along the equator: the west limb, east limb, antistellar point, and substellar point. 

A variety of internal structures are needed to explain the diversity of radii observed for hot Jupiters of similar masses \citep[e.g.,][]{2014arXiv1405.3752G,2017ApJ...844...94K}. All mechanisms that aim to explain the radius inflation in hot Jupiters invoke a higher entropy interior \citep{guillot-showman-2002}, including: ohmic dissipation \citep[e.g.,][]{2010ApJ...714L.238B}, downward energy flux via circulation \citep[e.g.,][]{2015ApJ...803..111G} or gravity waves \citep[e.g.,][]{2010ApJ...714....1A}, tidal heating \citep[e.g.,][]{2009ApJ...702.1413M}, increased IR opacities \citep[e.g.,][]{2007ApJ...661..502B}, inefficient heat transport in the interior \citep[e.g.,][]{chabrier-baraffe-2007}, and downward entropy mixing \citep{2017ApJ...841...30T}. We therefore consider two extreme cases for the interior of a given planet: the case of a high entropy interior, illustrated by the mechanism from \citet{2017ApJ...841...30T}, and a low entropy interior with T$_{\rm int} \sim 100$ K. 

 \begin{figure}[tbp]
\epsscale{1.17}
\plotone{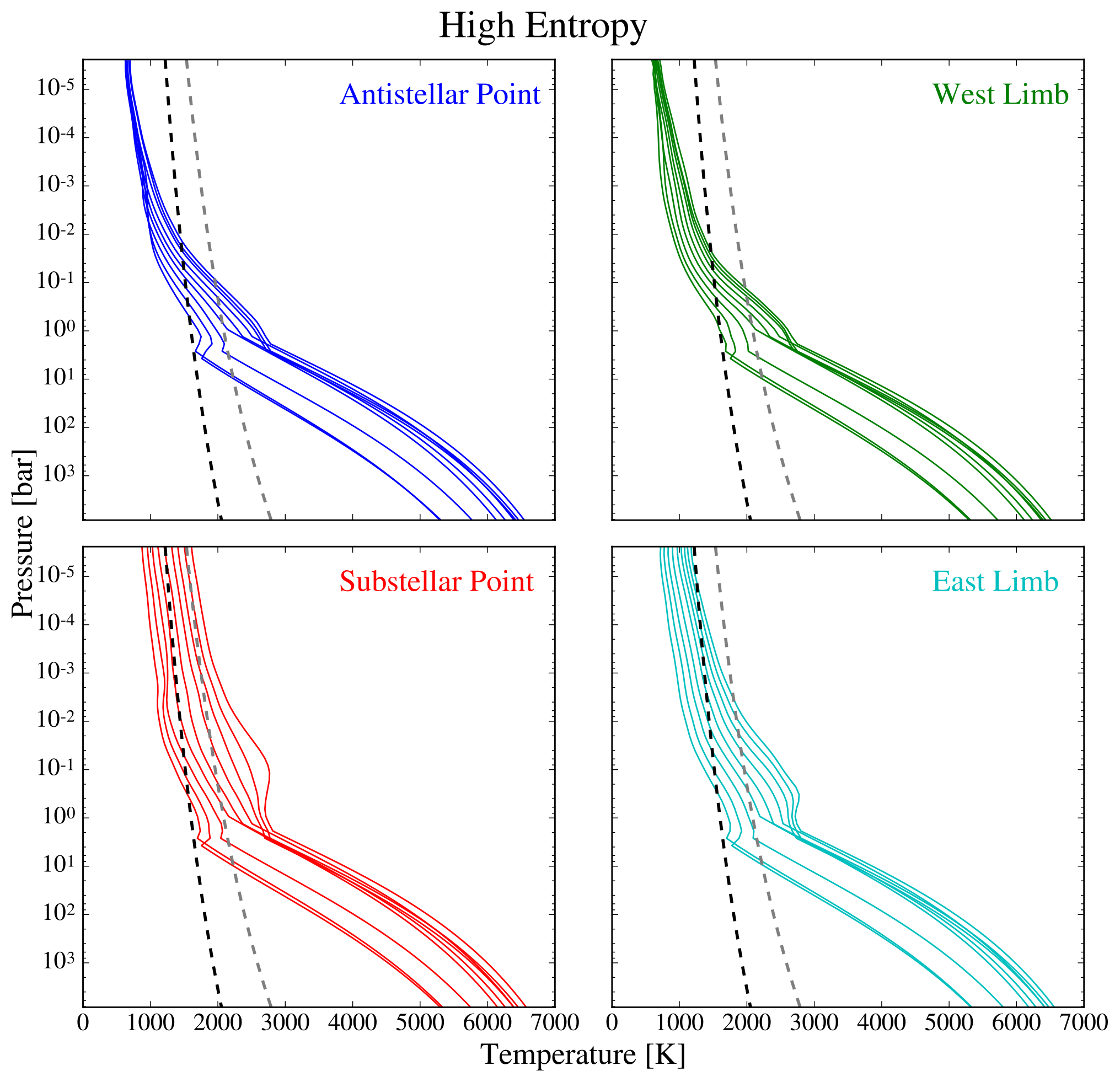}
\plotone{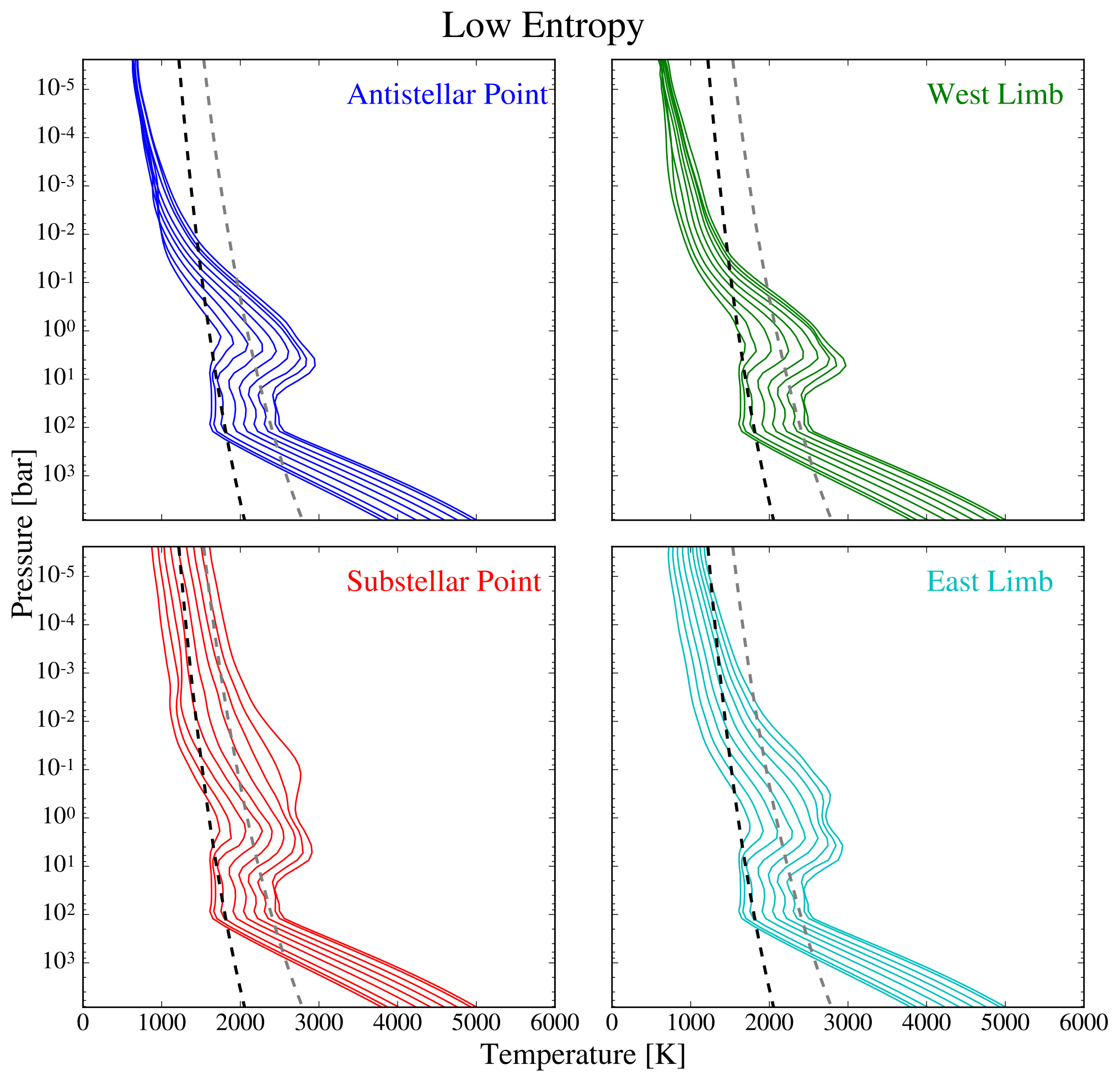}
\caption{Top Panel: High entropy interior pressure and temperature profiles for four representative locations in a hot Jupiter atmosphere. These profiles were created by combining a constant adiabat to the GCM output pressure and temperature profile below $\sim$ 3 bar. In each, the profile with the coolest equilibrium temperature (1300 K) is the leftmost line and profiles increase in equilibrium temperature in 100 K steps. The dashed lines shown correspond to the condensation curves of TiO$_2$ (gray) and MgSiO$_3$ (black). Bottom Panel: The same but for low entropy interior pressure and temperature profiles. These profiles were created by combining a constant adiabat to the base of the GCM output pressure and temperature profile at $\sim$ 100 bar.}
\end{figure}\label{he_prof}

The mechanism from \citet{2017ApJ...841...30T} relies on the advection of potential temperature to the interior of a planet triggered by non-uniform atmospheric heating. This allows us to use the temperature profile from the upper atmosphere to constrain the temperature at depth. Given this understanding, these two extreme cases correspond to two different efficiencies of entropy mixing in hot Jupiter atmospheres. In the high entropy case entropy mixing is efficient and the planet is inflated, with a hot interior; the opposite is true for the low entropy case. 

To create our full pressure and temperature profile for the high entropy case we therefore utilize the GCM profiles to roughly 3 bar of pressure---a point where the profiles at all representative locations converge. At this point the atmosphere is optically thick, such that assumptions made about the deep atmosphere will not change the resulting spectra. Below 3 bar, we assume that the planet has fully advected its potential temperature to the interior. The pressure and temperature profile of the planet can therefore be described by an adiabat below this point. Here we assume an adiabatic gradient of

\begin{equation}
\nabla_\text{ad} = 0.33-0.1(\text{T}/3000\;\text{K}) 
\end{equation}

\noindent for molecular hydrogen \citep[Equation 13]{2015A&A...574A..35P}. The resulting temperature profiles for the high entropy case are shown in the top panel of Figure \ref{he_prof}. 

We take the high entropy interior as the default case. Furthermore, we consider the complimentary case of a low entropy interior to investigate the physics of cold traps and to understand how differences in planet interiors can impact cloud properties.  

For the low entropy interior case, we use the full pressure and temperature profiles from the GCM. For P$> $3 bar, this solution is close to the initial condition; a 1D planet averaged model with T$_{\rm int} = 100$ K \citep[see][]{2015A&A...574A..35P}. We assume an adiabat below 100 bar, where the GCM profile ends. The resulting pressure and temperature profiles are shown in the bottom panel of Figure \ref{he_prof}. 

A notable feature of these profiles is the presence of an approximately isothermal region at roughly 10 bar in all profiles. As discussed in Section \ref{intuit}, the presence of an isothermal region can cause a supersaturation at two distinct points in the atmosphere. This region can therefore have an effect on the cloud properties and on the presence of a deep cold trap. Varying our choice of vertical resolution for both cases did not change the resultant cloud population. 

\begin{figure*}[tbp]
\epsscale{1.}
\plotone{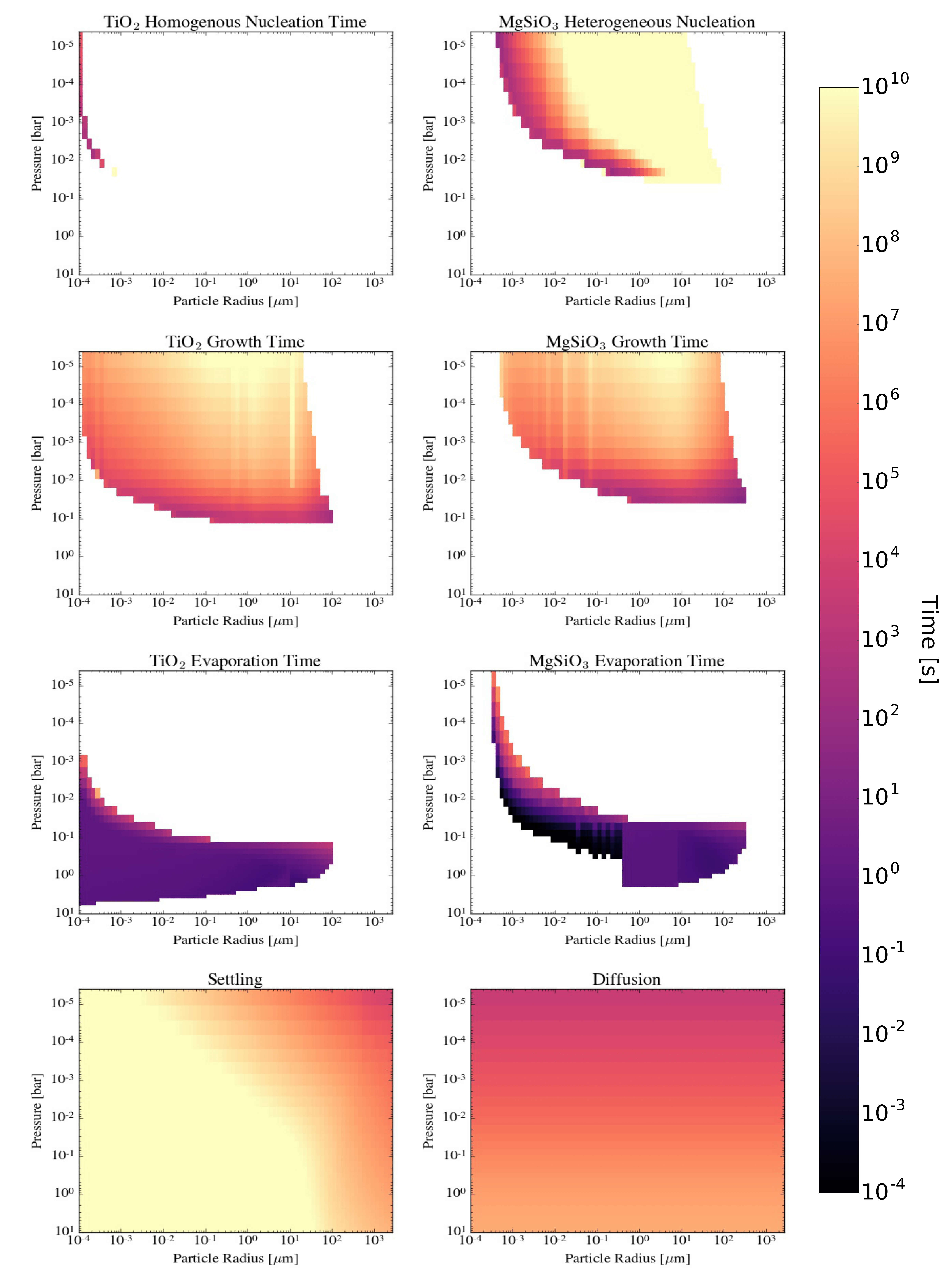}
\caption{The timescales of relevant microphysical and atmospheric dynamic processes. All processes are plotted as a function of the CARMA model grid in terms of particle radius and pressure. The white spaces are points in the atmosphere where either cloud particles are not present or they are not undergoing that process. The growth of TiO$_2$ and MgSiO$_3$ clouds, the heterogeneous nucleation of MgSiO$_3$, and the settling of particles occur relatively slowly. The homogeneous nucleation of TiO$_2$ and the diffusive vertical mixing occur more quickly. The evaporation of both species occurs rapidly when favorable. }
\end{figure*}\label{Timescales}

\subsubsection{Vertical Mixing}\label{vertmix}
The strength of vertical mixing in a planet plays an important role in determining the properties of the planetary atmosphere and its constituents, as discussed in Section \ref{transpo}. For simplicity, we adopt the one-dimensional parameterized K$_\text{zz}$ from \citep{2013A&A...558A..91P} for a canonical HD 209458b. This K$_\text{zz}$ takes the following form:

\begin{equation}
K_\text{zz} = \frac{5\times10^8}{\sqrt{P}} \text{ cm}^2\text{ s}^{-1},
\end{equation}

\noindent where $P$ is pressure in bar. 

This parameterization of K$_\text{zz}$ is derived from GCM modeling and is valid in the upper regions of a hot Jupiter atmosphere where the GCM pressure and temperature profile is used. In order to investigate the cloud properties in the deep atmosphere we set our K$_\text{zz}$ equal to a constant value of $5\times10^8$ cm$^2$ s$^{-1}$ below 3 bar. 

To test the sensitivity of our results to K$_\text{zz}$ we further vary the coefficient in the numerator as well as the constant value below 3 bar. We therefore additionally consider a K$_\text{zz}$ coefficient of $5\times10^7$ and $5\times10^9$ cm$^2$ s$^{-1}$.

\section{Timescales of Relevant Microphysical Processes}\label{timescales}

The processes of cloud microphysics depend sufficiently on the atmospheric parameters such that the timescales of these processes vary significantly with planetary properties. However, an understanding of the timescales of these processes can provide substantial insight into the resultant distribution of cloud particles. Before we present the detailed simulation results, we analyze the timescales of microphysical processes for a fiducial run of our hot Jupiter model: a high entropy interior hot Jupiter with an equilibrium temperature of 1700 K at the antistellar point. The processes that play an active role in governing the size distribution of cloud particles in our modeling are: the homogenous nucleation of TiO$_2$, the heterogeneous nucleation of MgSiO$_3$ on top of the TiO$_2$ CCN, the growth and evaporation of both MgSiO$_3$ and TiO$_2$, the settling of particles, and the diffusion of both gas and cloud particles. The timescales of these processes are shown in Figure \ref{Timescales} for our fiducial case. 

The nucleation, growth, and evaporation timescales are calculated using flux outputs from the CARMA model. Once the run reaches a steady state (for more details see Section \ref{results}) we determine the flux into (or out) of a given bin, time averaged over three months in model time, in units of cm$^{-3}$ s$^{-1}$. The number density in a given bin is then divided by these flux values to arrive at our estimated timescales. 

In all of our cases, cloud formation occurs above the point where the saturation vapor pressure is equal to the partial pressure of the species in the atmosphere (the point where the condensation curve crosses the pressure and temperature profile), known as the lifted condensation level (LCL) which can be a rough estimate of the cloud base level. This location varies in our modeling with the thermal structure of a given planet, with TiO$_2$ having a lower cloud base than MgSiO$_3$. For this fiducial case, the cloud base for TiO$_2$ is located at $3.4\times10^{-1}$ bar and the cloud base for MgSiO$_3$ is located at $7.2\times10^{-2}$ bar. 

In our model, gas diffuses from a well mixed interior into the upper atmosphere through vertical mixing. The timescale of this process can be approximated as the time that it takes to diffuse across an atmospheric scale height, i.e., $\tau_\text{diff} = H^2/\text{K}_\text{zz}$ where $H$ is the scale height. For the upper atmosphere above the cloud base it takes $10^3 - 10^5$ seconds for the gas to diffuse to an equilibrium state.  When the model is at equilibrium, the partial pressure of a given gas species closely follows its saturation vapor pressure curve. This is because the microphysical processes that deplete the gas are faster than gaseous diffusion. 

Once the gas has diffused above the cloud base, homogenous nucleation of TiO$_2$ cloud particles occurs. This nucleation takes roughly $10^3$ seconds, making it a moderately paced process. 

After small TiO$_2$ particles form via homogenous nucleation, these particles are able to grow by condensation or be heterogeneously nucleated upon by MgSiO$_3$. These particles can also evaporate, sediment, or be diffusionally lofted. When TiO$_2$ cloud particles evaporate, TiO$_2$ gas is released. The condensational growth of TiO$_2$ occurs slowly for most of the upper atmosphere ($\sim10^8$ seconds), but is significantly faster near the cloud base ($\sim10^3$ seconds). The evaporation of TiO$_2$ primarily occurs below the cloud base and for very small particles. This evaporation occurs relatively quickly, on timescales of $\sim 1$ second. 

The heterogeneous nucleation of MgSiO$_3$ onto TiO$_2$ occurs relatively slowly, particularly for particles larger than one micron. While heterogeneous nucleation happens the quickest for the smallest particles, these particles are also susceptible to evaporation, which occurs quickly for small particles throughout the cloud forming region (see Section \ref{ST}). The larger MgSiO$_3$ particles that form only evaporate below the MgSiO$_3$ cloud base where evaporation is rapid for particles of all sizes. When MgSiO$_3$ cloud particles evaporate, the component gases (e.g., Mg, SiO, H$_2$O) are released. The TiO$_2$ core is then able to evaporate into gaseous TiO$_2$ or survive as its own particle. TiO$_2$ particles are able to grow unimpeded unless they are nucleated on by MgSiO$_3$. Once a mantle of MgSiO$_3$ has formed only silicate condensation can occur.

Once MgSiO$_3$ has nucleated on a TiO$_2$ CCN, these clouds are also free to undergo microphysical and vertical transport processes. The condensational growth of MgSiO$_3$ occurs at roughly the same pace as the growth of TiO$_2$ and is again fastest at the cloud base. 

\begin{figure}[tbp]
\epsscale{1.3}
\plotone{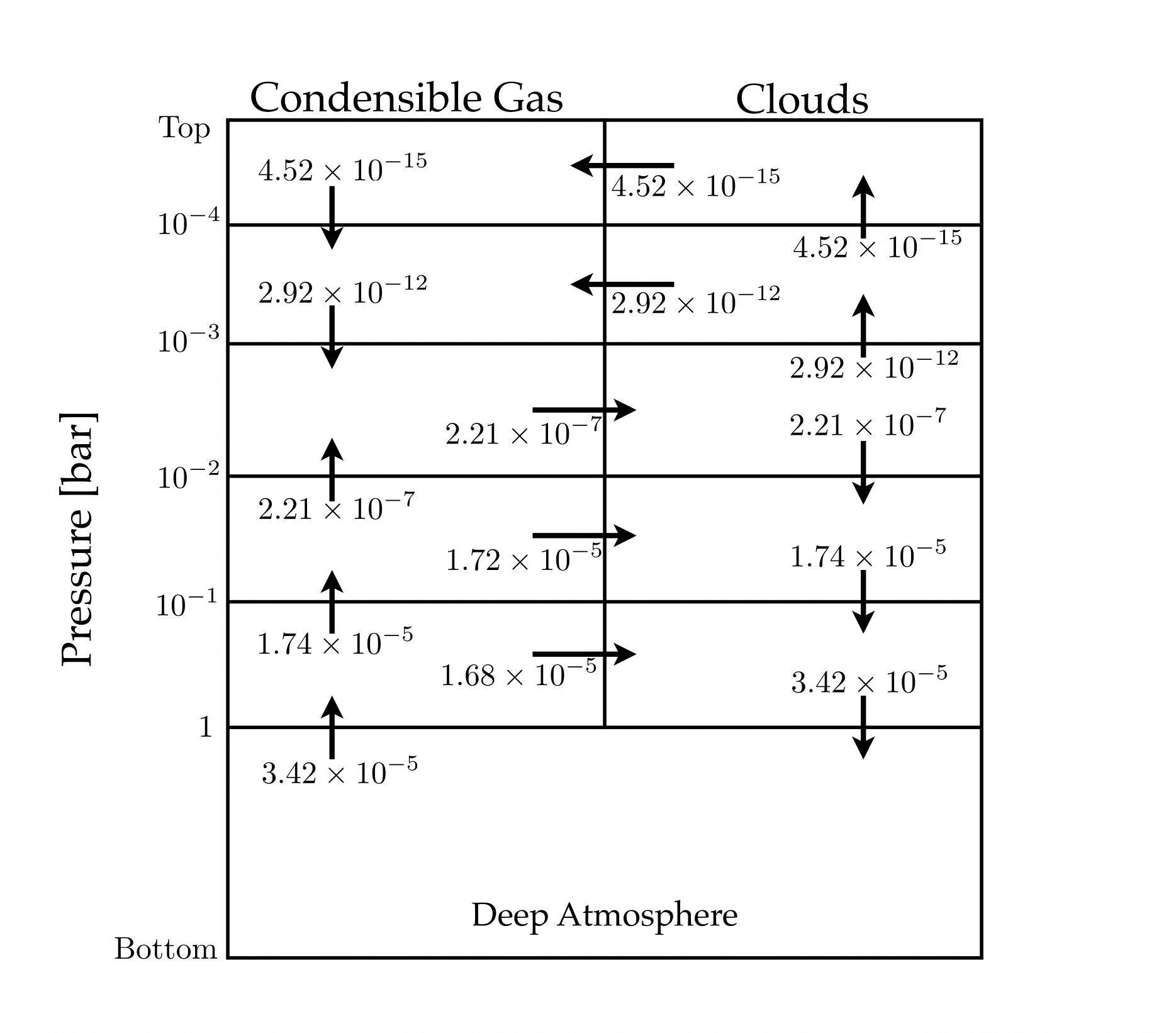}
\caption{Condensible species flux flow (in units of g cm$^{-2}$ s$^{-1}$) for a hot Jupiter with T$_\text{eq} = 1700$ K at the antistellar point.}
\end{figure}\label{massbalance}

Gravitational settling further acts on all cloud particles. We approximate the settling timescale as the time that it takes for a particle to settle through an atmospheric scale height, i.e., $\tau_\text{settle} = H/v_\text{fall}$ where $H$ is the scale height and $v_\text{fall}$ is the settling velocity of the particle calculated in CARMA \citep[see][Appendix A]{2018arXiv180206241G}. Particle settling happens at a relatively slow pace, particularly for particles smaller than $\sim$ 10 microns, for which settling across a scale height takes $10^9-10^{10}$ seconds. This timescale gradually transitions to faster times, however, and is noticeably more efficient for particles larger than 10 microns, which can settle in $\sim 10^5$ seconds. Given our fiducial diffusivity profile, diffusive transport dominates settling for nearly all relevant particle sizes.

While these timescales vary with atmospheric location and particle size, they are roughly ordered in magnitude as described in Equation \ref{timey}.
\begin{multline}\label{timey}
\tau_\text{evap, MgSiO$_3$} \sim \tau_\text{evap, TiO$_2$} << \tau_\text{nuc, TiO$_2$} \sim \tau_\text{diff} \\< \tau_\text{gr, MgSiO$_3$}  \sim \tau_\text{gr, TiO$_2$}  \sim \tau_\text{setl} \sim \tau_\text{nuc, MgSiO$_3$}
\end{multline}

These timescales change throughout the atmosphere such that just above the cloud base, cloud particles are dominated by condensational growth, whereas higher in the atmosphere they are dominated by nucleation and vertical transport. 

A picture of the mass balance in the atmosphere for this fiducial case is shown in Figure \ref{massbalance}. Most of the cloud formation processes occur near the cloud base and at pressures higher than 10$^{-3}$ bar. Below $\sim 10^{-3}$ bar particles preferentially experience settling, while above this point particles are more likely to be lofted upwards via vertical mixing.

\begin{figure}[tbp]
\epsscale{2.4}
\plottwo{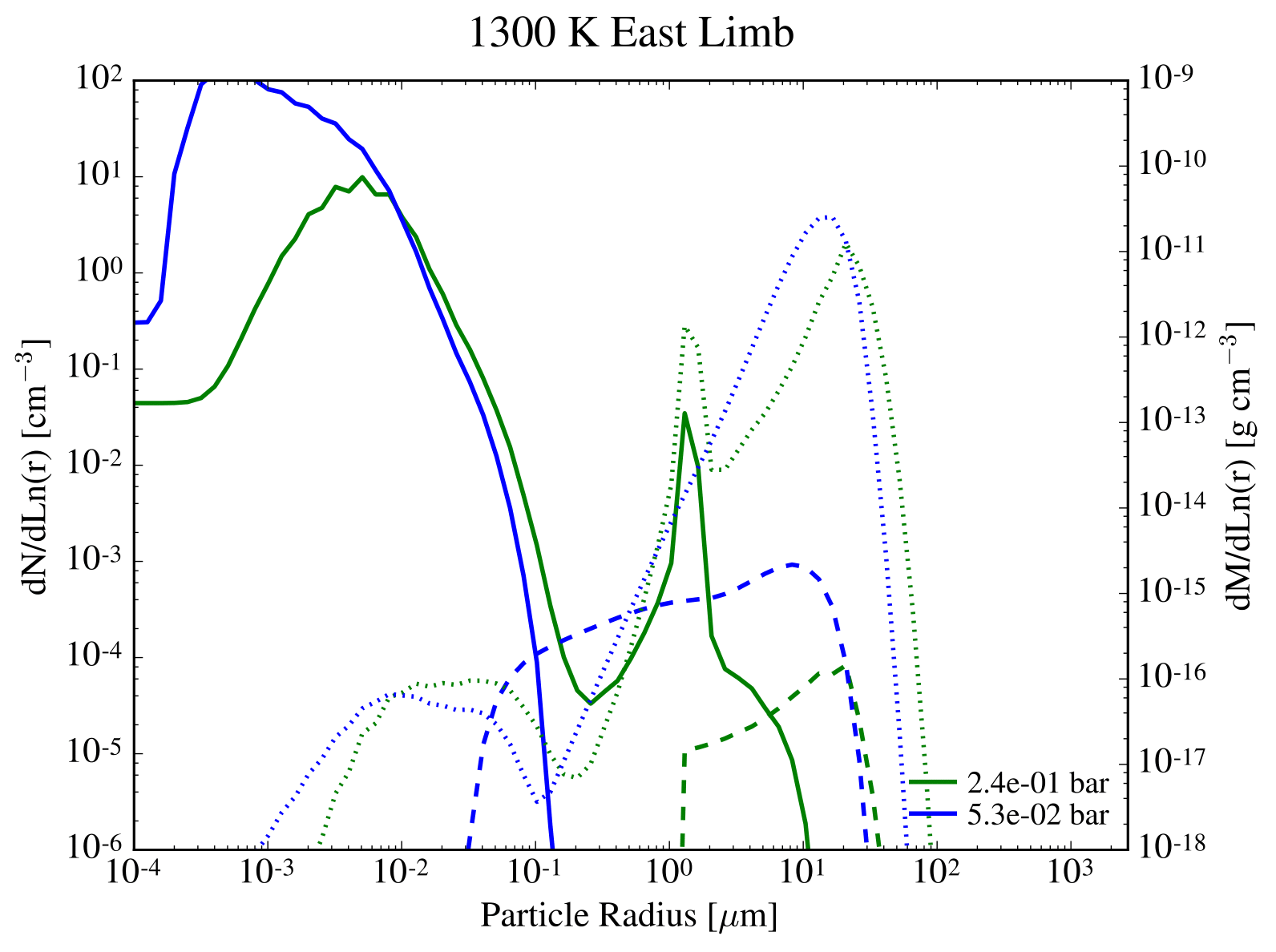}{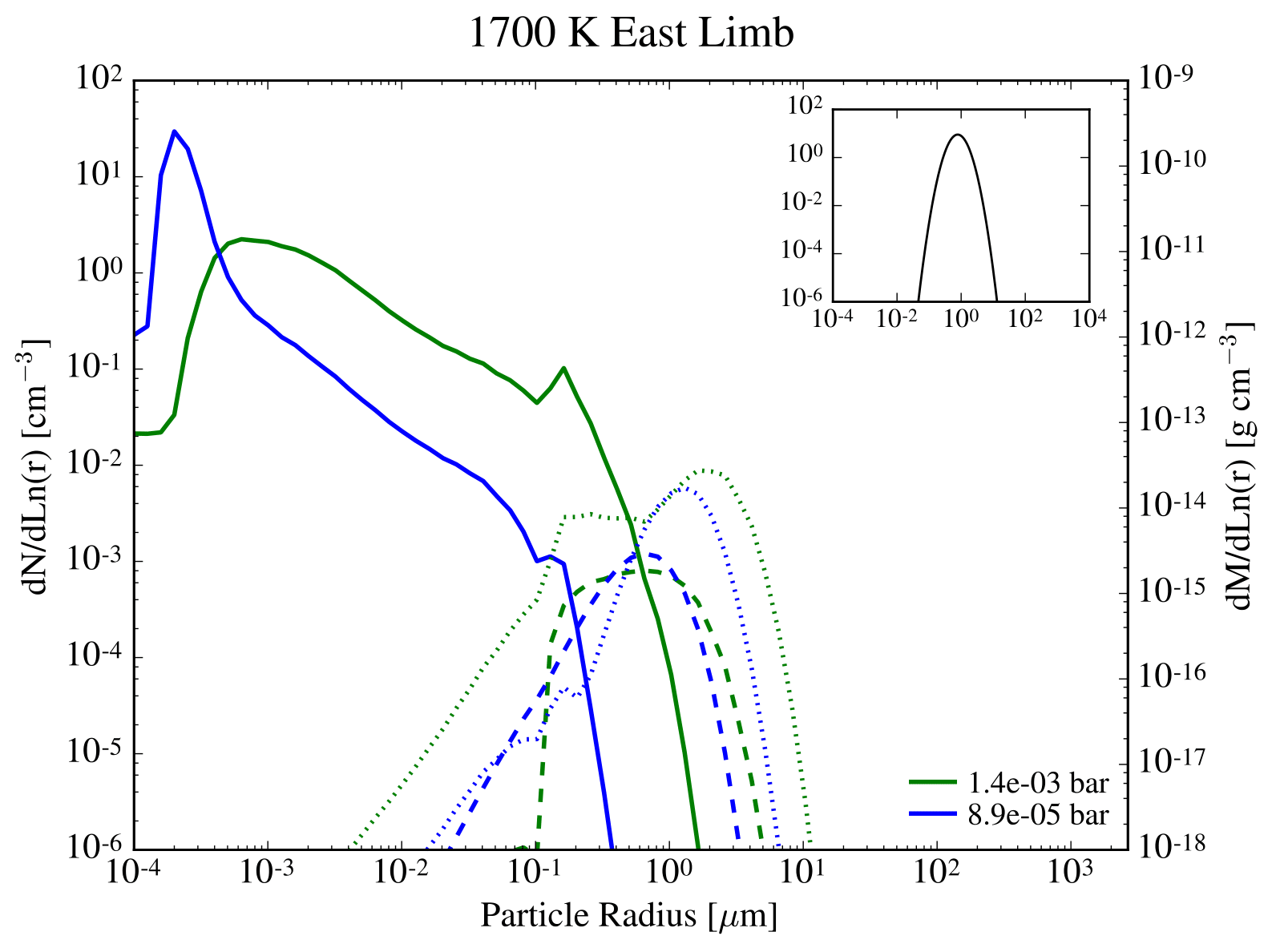}
\caption{Cloud particle size distributions in terms of number density (solid lines for titanium clouds and dashed lines for silicate clouds) and mass density (dotted lines for titanium and silicate clouds added together) for two representative hot Jupiters. Size distributions are plotted for a specific pressure in the atmosphere as indicated in the legend. For the 1700 K case the inset plot depicts a standard log-normal size distribution. In all cases the cloud particle size distribution does not follow a smooth log-normal profile.}
\end{figure}\label{notlog}

\section{Simulation Results}\label{results}
We calculate the cloud particle size distributions, the total cloud mass, and the vertical distribution of cloud particles for a grid of 9 Jupiter-size tidally locked planets orbiting a solar-type star with equilibrium temperatures ranging from 1300 K to 2100 K. We sample the atmosphere at four representative locations along the equator: the antistellar point, substellar point, east limb, and west limb. We further consider two representative cases for these planets' interiors: high entropy and low entropy. A comprehensive discussion of our model grid can be found in Section \ref{planet}.

In the following sections we discuss trends that are apparent in our results when time averaged over the last three Earth years of a thirty year run in model time. Our models arrive at a steady state solution rather than a true equilibrium \citep[see][]{gao2018} where we define our steady state as stable oscillations around a mean value as is seen in many 1D cloud formation models \citep{2003Icar..162...94B}. These oscillations occur on roughly Earth year timescales. This is suggestive of some intrinsic variability, though we leave further discussion for future work. In the following, we will mainly adopt the high entropy simulations as the nominal cases to discuss our findings, while the low entropy cases are merely used to test the effects of a deep cold trap.

 \begin{figure*}[tbp]
\epsscale{1.15}
\plotone{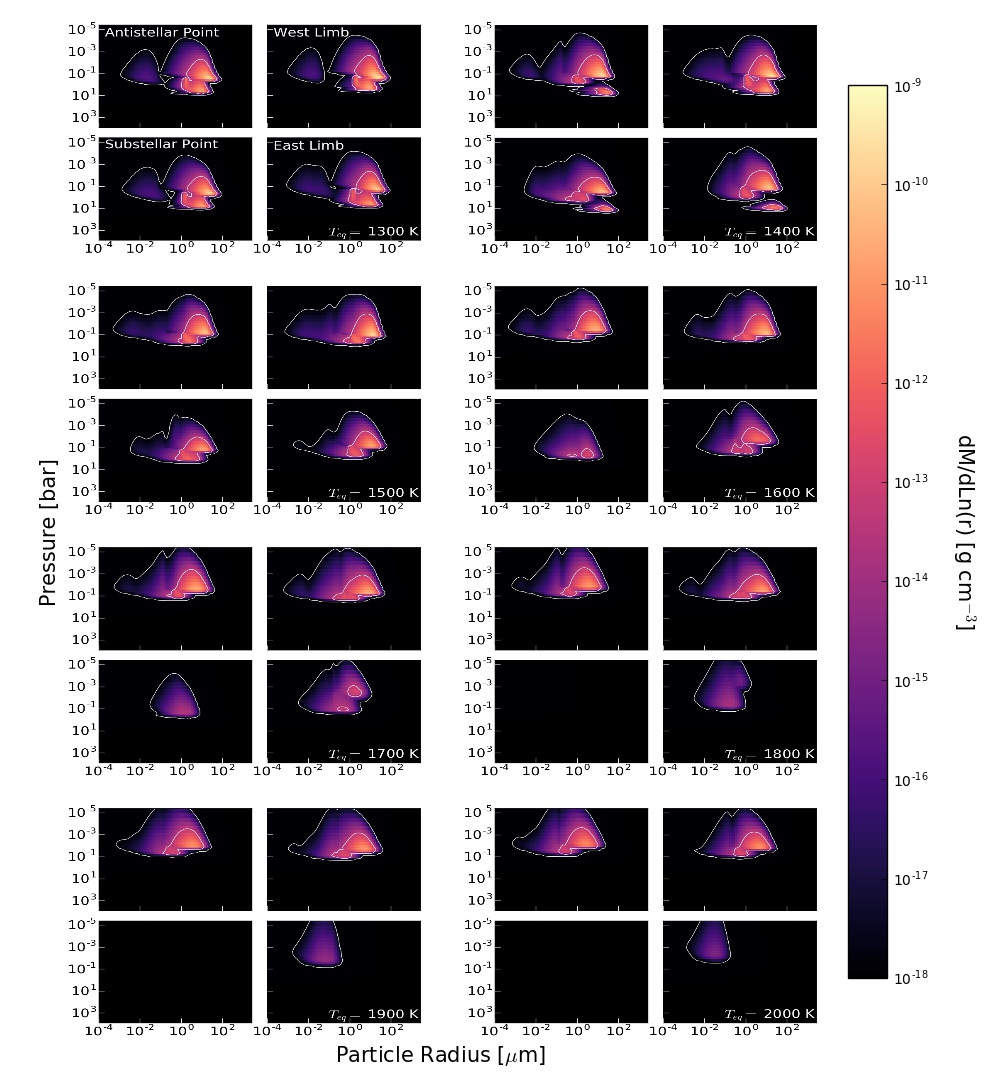}
\caption{Vertical cloud particle size distributions for the high entropy interior case in terms of mass density ($dM/dLn(r)$). Both TiO$_2$ clouds and MgSiO$_3$ clouds are plotted using the same colormap. All plots are made using a log-scale. The clouds appear vertically extended while the majority of the mass is close to the base of the cloud deck. The contours correspond to the range in the colorbar divided into 3 even sections in log-space. There are distinct trends in cloud properties with equilibrium temperature and planet location. The 2100 K equilibrium temperature case is excluded from this plot as the resultant size distributions are very similar to those from the 2000 K case. }
\end{figure*}\label{he1}

\subsection{Cloud Particle Size Distributions}
The resultant cloud particle size distributions in our grid are not log-normal and are instead bimodal, broad, or irregular in shape. Figure \ref{notlog} shows typical distributions for two representative equilibrium temperatures at two representative pressures in the atmosphere. It is important to note that the particle size distributions can vary significantly with altitude.

The silicate clouds are typically distributed broadly, sometimes without a distinct peak. The distribution of silicate cloud particles has a distinct peak closer to the cloud base where growth is efficient until it is limited by particle settling. The distribution has an indistinct peak when growth is less efficient and particles of nearly all sizes in the distribution can persist until they are limited by settling. Furthermore, the silicate clouds are frequently distributed asymmetrically such that the distribution skews towards smaller particles. 

The titanium clouds frequently follow a bimodal distribution with a peak at small radii (the nucleation mode) and another peak at intermediate radii corresponding to the particles that are able to overcome the Kelvin effect and grow to a larger size (the growth mode). The first peak at smaller radii is typically broad while the second peak at larger radii is narrow. At altitudes sufficiently above the cloud base only the nucleation mode is present in a broad distribution.

The CCN size on which silicate clouds can efficiently heterogeneously nucleate is approximately indicated by the size at which TiO$_2$ particle number densities drop below those of the silicate cloud particles. The existence of an optimal CCN size is due to the Kelvin effect, as smaller CCN are difficult to nucleate on without quickly undergoing evaporation while larger CCN are not as numerous. 

For the case of the high entropy planetary interior, the cloud particle distributions in terms of mass density ($dM/dLn(r)$) are shown in Figure \ref{he1}. Note that clouds are only present in the upper atmosphere in these cases. Here both the titanium and silicate cloud particles are plotted using the same colormap. The population of titanium cloud particles ranges in radius from 10$^{-1}$ to 1 $\mu$m and is typically smaller than the population of silicate cloud particles, which range in radius from 10 to 50 $\mu$m. 

When silicate clouds form in abundance, the titanium clouds form in two populations: below the silicate cloud base and above it. The titanium clouds that form below the silicate cloud base tend to grow larger in size than those that form above it as their growth is not limited by the heterogeneous nucleation of silicate clouds. 

Titanium cloud particles, if they form, are typically abundant throughout the upper regions of the atmosphere, while silicate cloud particles are confined closer to their cloud base. This is shown in Figure \ref{vert_sd} for the 1300 K hot Jupiter at the antistellar point. In Figure \ref{vert_sd} the titanium cloud particles are abundant from above 10$^{-1}$ bar to the top of the atmosphere while the silicate cloud particles are abundant closer to their cloud base and extend to roughly 10$^{-3}$ bar. This general trend is found for all cases where both clouds form.

\subsection{The Effects of Local Temperature and Equilibrium Temperature}\label{HE_sec}

 \begin{figure}[tbp]
\epsscale{1.2}
\plotone{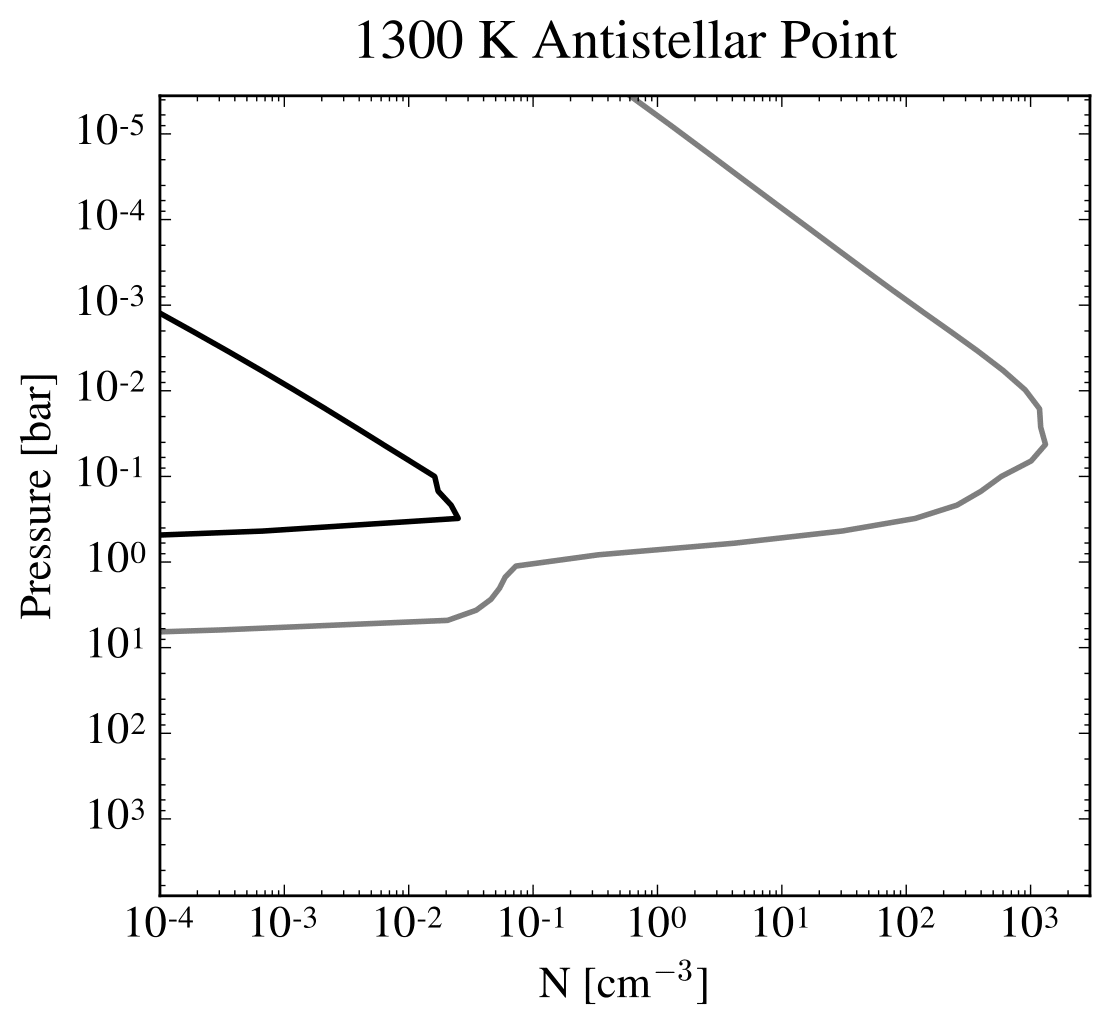}
\caption{Total number densities as a function of pressure in the atmosphere of a 1300 K hot Jupiter at its antistellar point for titanium (gray) and silicate (black) cloud particles. The titanium cloud particles are abundant from above 10$^{-1}$ bar to the top of the atmosphere. The silicate cloud particles are abundant closer to their cloud base and extend to roughly 10$^{-3}$ bar. }
\end{figure}\label{vert_sd}

The formation of clouds occurs at all four representative locations along the equator for planets with T$_{\rm eq} < 1800$ K. Planets with equilibrium temperatures greater than or equal to 1800 K have clear atmospheres (in terms of titanium and silicate clouds) at the substellar point, as the local temperature profile becomes too hot for cloud formation to occur. With increasing equilibrium temperature, the cloud base moves towards the upper atmosphere and the cloud cover becomes increasingly inhomogeneous as a function of longitude with the west limb and antistellar point being preferentially cloudy. Cloud particles located on hotter regions of the planet (the east limb and substellar point) tend to be smaller than the cloud particles present at cooler locations. This effect is due to the increase in temperature at the east limb and substellar point. The temperature increase changes the saturation vapor pressure leading to lower supersaturations. The lower supersaturations lead to limited growth and smaller mean particle sizes. This effect is particularly strong for planets with high equilibrium temperatures where the east limb and substellar points have particularly high temperatures. 

In some locations there exists only a relatively small population of titanium cloud particles, with no silicate clouds, while both clouds are abundant in other locations. For example, for equilibrium temperatures greater than or equal to 1900 K,  the east limbs only have a significant population of titanium clouds. The antistellar points and west limbs, however, have both titanium and silicate clouds for all equilibrium temperatures in our grid. This is a temperature effect as there are specific regions of temperature space for which TiO$_2$ reaches a supersaturation and can form clouds while it is too hot for MgSiO$_3$ cloud particles to form. Furthermore, the east limb and substellar points experience more dramatic increases in temperature with increased equilibrium temperature as compared to the west limb and antistellar point.

\begin{figure}[tbp]
\epsscale{1.2}
\plotone{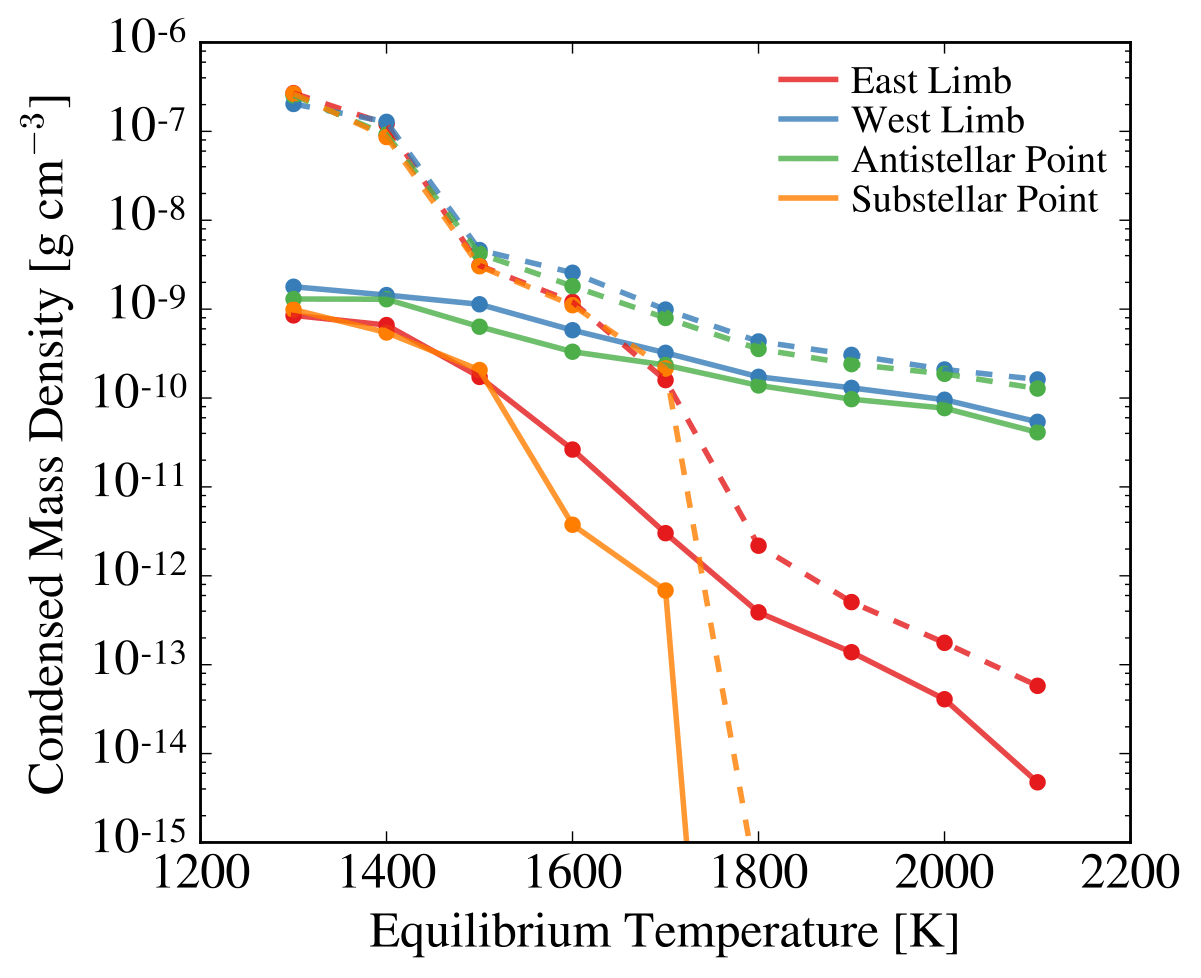}
\caption{Total condensed mass density as a function of equilibrium temperature for four representative planetary locations for the case of a high entropy interior (solid lines) and low entropy interior (dashed lines, see Section \ref{low}). All locations show a marked decrease in condensed mass density as a function of equilibrium temperature, with the trend being more pronounced for the east limb and substellar point.}
\end{figure}\label{he_masstrend}

The presence of a small local thermal inversion in the 1300 and 1400 K case (see Figure \ref{he_prof}) has an impact on the vertical locations of the cloud populations, such that there are two small and distinct cloud layers. This occurs because there are two locations in the atmosphere that reach a supersaturation, separated by a small region of pressure space that is too hot for a supersaturation to be achieved. However, this primarily affects the deep population of titanium clouds without strongly affecting the overall cloud distribution.  

There is a decrease in cloud mass density with increasing equilibrium temperature across all sampled regions of the planet, shown in Figure \ref{he_masstrend}. This is because an increase in temperature reduces the supersaturation for a given condensate partial pressure, resulting in less gas condensing. For nearly all cases, the west limb and the antistellar point form the same density of cloud particles to within an order of magnitude as their temperature profiles are also quite similar. The cloud particle size distribution in these locations differs subtly, however, with the west limb preferentially forming larger cloud particles in a slightly narrower distribution. This subtle change is due to the west limb having slightly cooler temperatures in the cloud forming region of the atmosphere, leading to an increased supersaturation and supply of condensible gas which causes increased particle growth.

The hotter regions of a hot Jupiter's atmosphere (the east limb and substellar point) show a more dramatic dependence on equilibrium temperature, as shown by the steeper slope in Figure \ref{he_masstrend}. At equilibrium temperatures lower than 1500 K, the east limb and substellar point also form roughly equal densities of cloud particles. The relatively flat slope for the antistellar point and west limb in Figure \ref{he_masstrend} indicates that cloud properties in the cooler regions of hot Jupiters may be relatively unaffected by increasing equilibrium temperatures, while hotter regions see much more dramatic changes leading to limited particle growth. 

\begin{figure}[tbp]
\epsscale{1.25}
\plotone{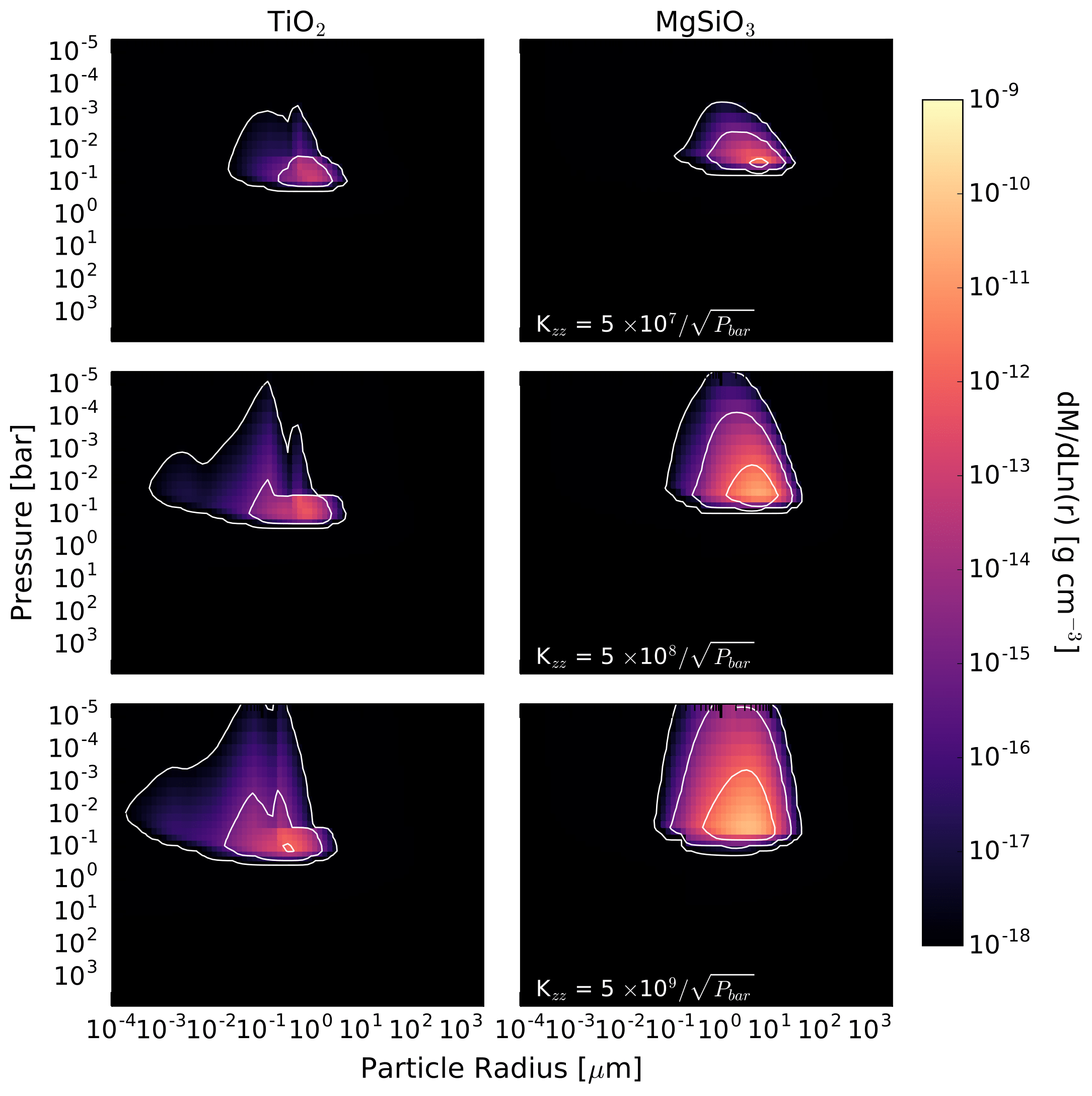}
\caption{Vertical cloud particle size distributions for a 1700 K hot Jupiter at the antistellar point for the case of a high entropy interior as a function of atmospheric vertical mixing: low (top), fiducial (middle), and high (bottom). TiO$_2$ clouds (left) and MgSiO$_3$ clouds (right) are plotted separately. There is an increase in total cloud mass and differences in the properties of the cloud particle size distribution with increased vertical mixing.}
\end{figure}\label{kzz_vert}

\subsection{The Influence of Vertical Mixing on Cloud Properties}
We choose the 1700 K hot Jupiter at the antistellar point as a fiducial case to determine the effect of atmospheric mixing on the cloud particle size distribution. To understand the effect that vertical mixing has on the distribution of cloud particles, we vary our input K$_\text{zz}$ by an order of magnitude---both smaller and larger. The distributions are shown in Figure \ref{kzz_vert}, where we plot the titanium and silicate clouds separately. 

When the atmospheric vertical mixing is reduced, the total cloud mass and vertical extent of both cloud particle populations are significantly smaller than in our nominal case. In particular, there is a decreased number of small titanium cloud particles. As atmospheric vertical mixing is increased, there is an increased population of both titanium and silicate clouds. With increased vertical mixing, the vertical extent of the cloud particle populations increases slightly while the mean particle size decreases slightly compared to our fiducial case. This is due to an increased production of particles leading to greater number of particles vying for the gas with which to grow, leading to on-average smaller particles. This effect is subtle, however, as the increased vertical mixing also increases the available supply of condensible gas.

The enhanced vertical extent of the cloud population is primarily due to this increased supply of condensible gas to the cloud forming region of the atmosphere. The increased supply of gas leads to more growth and extends the region of rapid growth further above the cloud base. This leads to both an increase in cloud mass and vertical extent. There is also the secondary effect that particles are lofted higher in the atmosphere further extending the region of abundant cloud particles.

The total mass density of titanium and silicate clouds is strongly correlated with the amount of vertical mixing in the atmosphere. This is shown in Figure \ref{kzz_mass}, where the total cloud mass density increases substantially with increased mixing. 

\begin{figure}[tbp]
\epsscale{1.}
\plotone{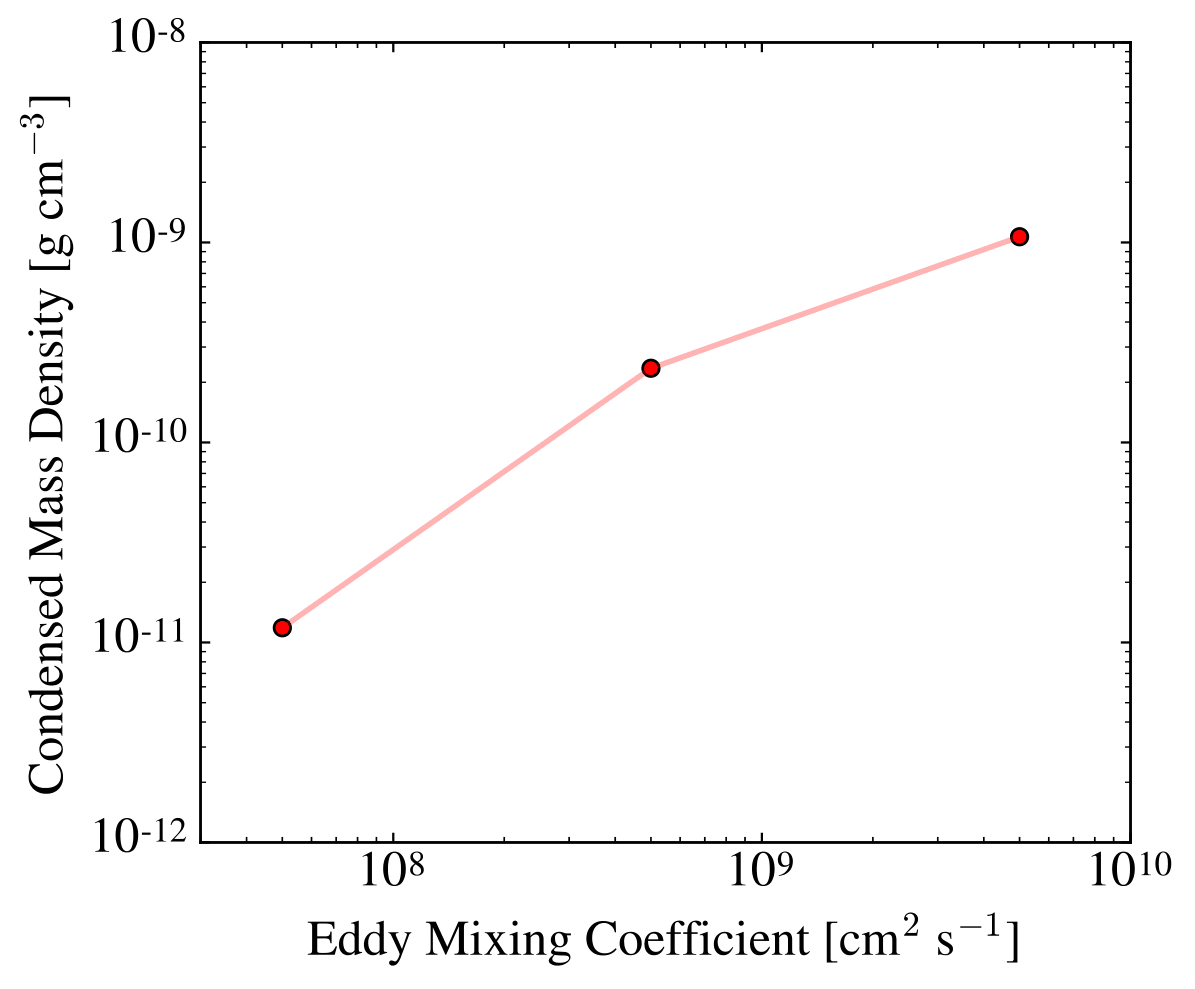}
\caption{Total condensed mass density as a function of vertical mixing for a 1700 K hot Jupiter at the antistellar point for the case of a high entropy interior. There is a marked increase in total condensed cloud mass with increased vertical mixing.}
\end{figure}\label{kzz_mass}

\begin{figure*}[tbp]
\epsscale{1.}
\plotone{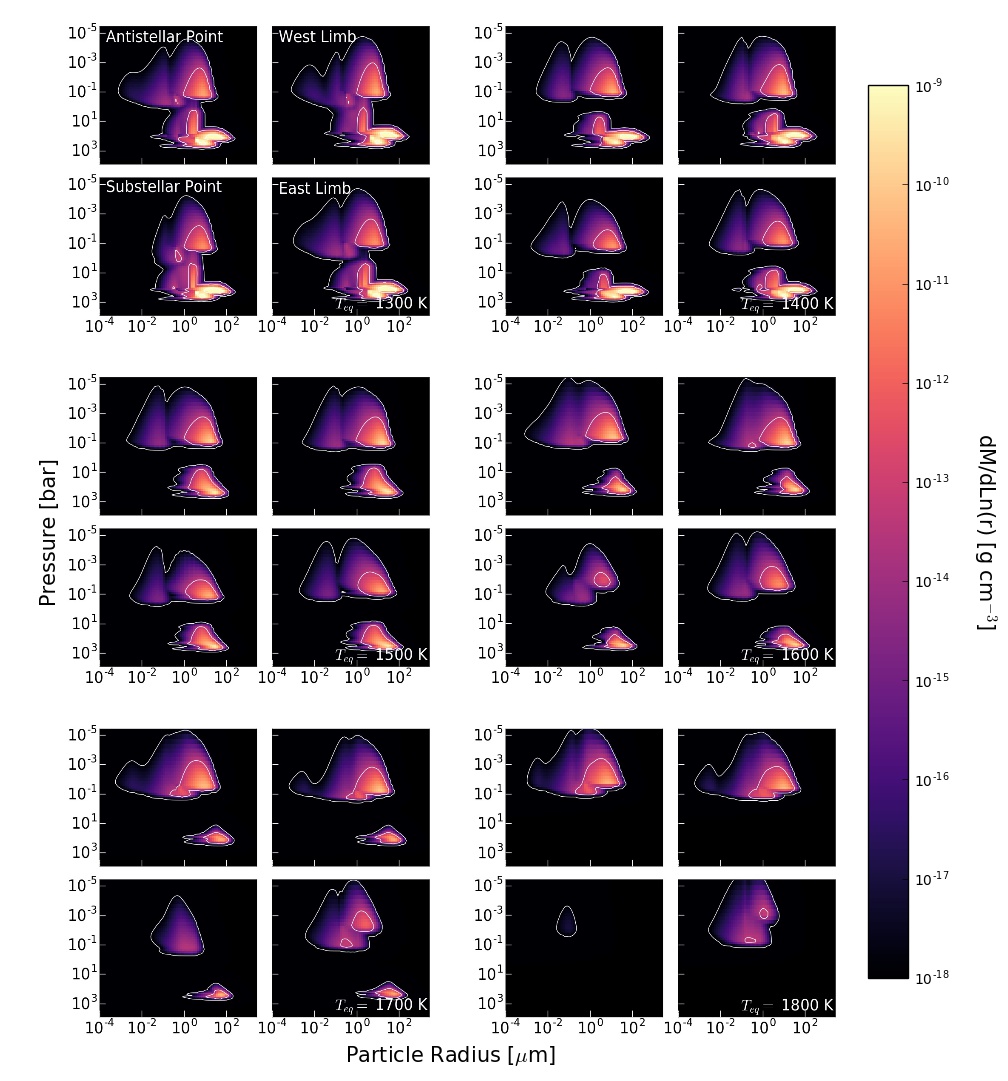}
\caption{Same as Figure \ref{he1}, but for the low entropy interior case with an emphasis on equilibrium temperatures that have a deep cold trap. }
\end{figure*}\label{le1}

\subsection{Low Entropy Temperature Profile and the Presence of a Deep Cold Trap}\label{low}
We focus on the resulting cloud particle distributions for the low entropy cases in which a deep cold trap is present in the lower atmosphere---resulting in marked differences from the high entropy interior cases. This is true in our grid for hot Jupiters with equilibrium temperatures lower than 1800 K. For planets with equilibrium temperatures of 1800 K or higher the cloud particle distributions are very similar to those shown in the high entropy case in Section \ref{HE_sec}. The cloud particle distributions in terms of mass density are shown in Figure \ref{le1}. Again, both the titanium and silicate cloud particles are plotted using the same colormap. 

The presence of an isothermal region in the pressure and temperature profile leads to the formation of two cloud populations that are spatially separated in the atmospheres of hot Jupiters (see bottom panel of Figure \ref{he_prof}) for one or both of our cloud species for planets with temperatures below 1800 K. These populations exist because the isothermal layer reduces the temperature at depth, leading to the existence of two regions in the atmosphere where supersaturation can be achieved, separated by a region that is too hot for clouds to form (see Section \ref{coldcold}). The one exception is the 1300 K case, where the temperatures are low enough to allow the two cloud populations to merge. 

The first population of clouds is present in the deep atmosphere, at around 100 bar. This lower cloud deck is comprised of large cloud particles, with both titanium and silicate cloud particles growing to tens or hundreds of microns in size due to a large supply of gas at depth. This population of clouds varies in vertical extent with equilibrium temperature. At cooler temperatures cloud particles extend throughout most of the atmosphere while at hotter temperatures the lower cloud deck is confined to the deep atmosphere. This is because the layer of the atmosphere in which it is too hot for clouds to form becomes larger with increased equilibrium temperature (see Figure \ref{he_prof}). 

We refer to the lower population of clouds in this atmosphere as a deep cold trap (see Section \ref{coldcold}). This deep cold trap theoretically limits cloud formation in the upper atmosphere; however, for all of the planets in our grid with a deep cold trap, cloud formation in the upper atmosphere appears only subtly affected. This is because atmospheric mixing is strong enough to supply the upper atmosphere with sufficient gas for abundant cloud formation. We further discuss the efficiency of the deep cold trap in altering atmospheric observables in Section \ref{obs}.

Increasing the equilibrium temperature of a planet decreases the total amount of cloud mass, as shown in Figure \ref{he_masstrend}, following the same reasoning as for the high entropy case. In contrast to those cases, however, there is a much larger cloud particle mass density for atmospheres with a deep cold trap (T$_\text{eq} < 1800$ K) because this deep reservoir adds mass without substantially limiting supply to the upper atmosphere. Planets with equilibrium temperatures less than 1800 K form a nearly homogenous layer of clouds in the deep atmosphere such that the total condensed mass density is the same across all four planetary locations. 

For equilibrium temperatures greater than 1800 K, where no deep cold trap is present and supersaturation is only achieved in the the upper atmosphere, the four locations again differ in cloud particle mass density. In particular, the west limb and antistellar point have very similar cloud particle mass densities while the east limb and substellar point show a stronger dependence on equilibrium temperature as seen in the high entropy case.

\subsection{Comparison to Other Modeling Approaches}\label{compmodel}
Our modeling framework differs considerably from models that rely on equilibrium cloud condensation or on grain chemistry. Here we summarize the similarities and differences between our study and previously published work. We note, however, that any differences in assumed temperature profile could also result in differences between the studies, in addition to the differences caused by different modeling frameworks.

The modeling framework described in \citet{ackerman-marley-2001} assumed that clouds are not present below the cloud base. Indeed, none of our simulations produce abundant cloud particles below the cloud base, as evaporation occurs quickly. This finding indicates that this assumption is likely valid to first order and that the cloud base is thermodynamically controlled. 

Previous modeling work done by \citet{2015A&A...580A..12L} for HD 189733b found that silicate clouds are the main component of the total condensible inventory. While we do not consider a comprehensive list of condensible species, our simulations find that silicate clouds do dominate titanium clouds in terms of mass in most cases when both species are present. 

Follow-up work for HD 189733b by \citet{2016A&A...594A..48L} found that the hottest regions of the atmosphere along the equator are populated by the smallest cloud particle grains. Our modeling also uncovers this trend, although the effect is sometimes subtle. Furthermore, the mean particle sizes of our clouds, particularly near the cloud base, are very similar to those derived in \citet{2016A&A...594A..48L}. Unlike the modeling done in \citet{2016A&A...594A..48L}, we do not consider horizontal mixing which could work to smooth inhomogeneities in cloud coverage with longitudinal location.

Our ability to predict fully resolved size distributions allows us to test common assumptions. \citet{ackerman-marley-2001} assume a log-normal distribution of cloud particles, and grain chemistry modeling as used in \citet{2015A&A...580A..12L,2016A&A...594A..48L} uses the moment method to derive four governing parameters of a smooth particle size distribution. Our results do not support these assumptions; we instead find a varying cloud particle size distribution that is frequently bimodal or irregular in shape due to both cloud composition and formation mechanisms.

In contrast to some of the results in \citet{2017ApJ...847...32L}, we do not find that considering coagulation in our modeling has a significant effect on our derived cloud particle distribution. This result is unsurprising, however, as our work focuses on condensational clouds with much lower number densities than the photochemical hazes considered in their work. The maximum particle number densities we encounter in our results are $\sim 10^2$ cm$^{-3}$ for the high entropy interior cases, while \citet{2017ApJ...847...32L} consider number densities greater than $10^4$ cm$^{-3}$.

\section{Observational Implications}\label{obs}
We now discuss the observational implications of our derived cloud particle size distributions in detail. In the following sections we only discuss radiative properties of the cloud particles themselves given our derived cloud particle size distributions and do not consider the opacities of the background gases. We do so as a means to clearly understand how the radiative properties of clouds depend on planetary properties and their underlying size distribution. 

\begin{figure*}[tbp]
\epsscale{1.}
\plotone{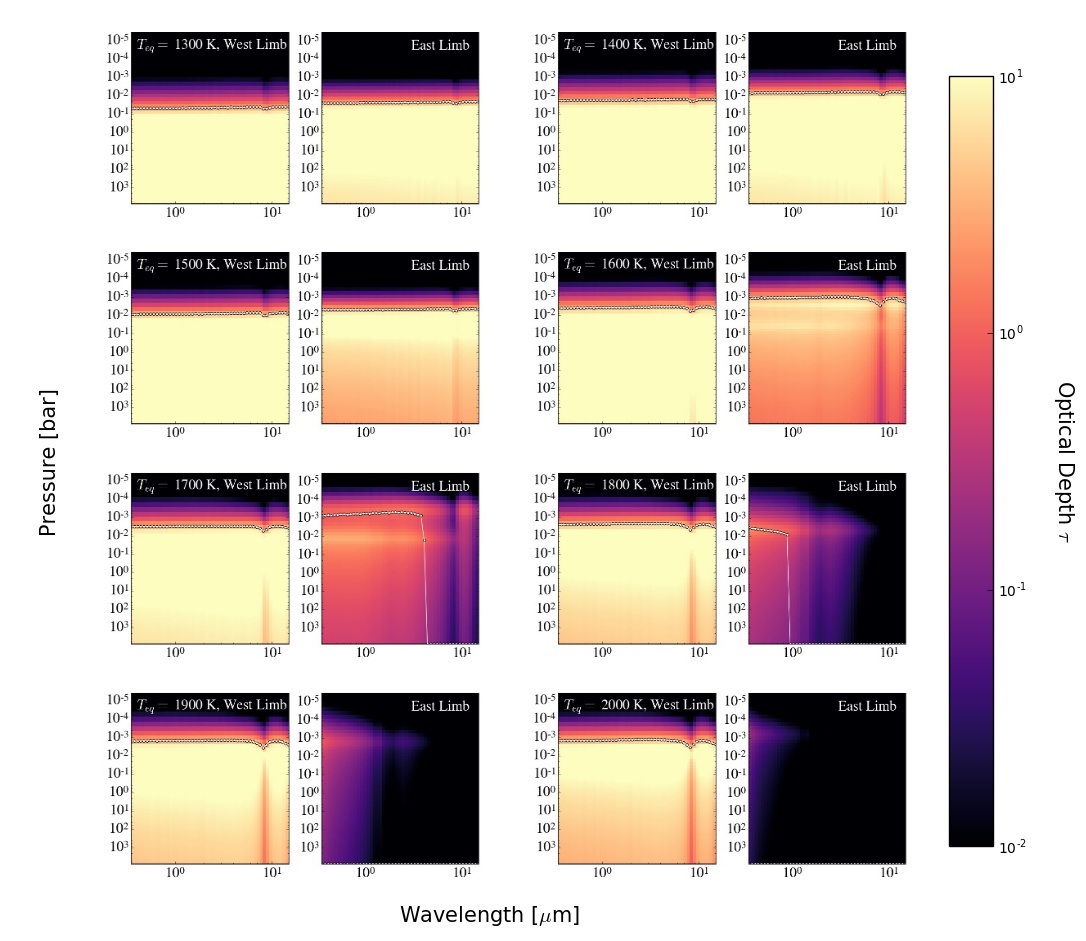}
\caption{Cloud transmission opacities for the case of the high entropy interior. The white dotted line represents the point in the atmosphere where the clouds become opaque---the ``opaque cloud level". There are noticeable hemispheric differences between the east and west limbs for hotter planets. The opaque cloud level is at roughly the same location for a range of wavelengths and equilibrium temperatures. }
\end{figure*}\label{trans_he}

First, we calculate cloud opacities and scattering properties in transmission and emission observational viewing geometries. We focus our discussion in part on the differences in cloud radiative properties between the high and low entropy cases as well as longitudinal differences in a planet's atmosphere. We also discuss trends in cloud opacity with equilibrium temperature. Second, we investigate the impact of using a full cloud particle size distribution in opacity calculations. In the following sections we focus solely on cloud opacities and other specific properties of our derived cloud populations. 

\begin{figure}[tbp]
\epsscale{.9}
\plotone{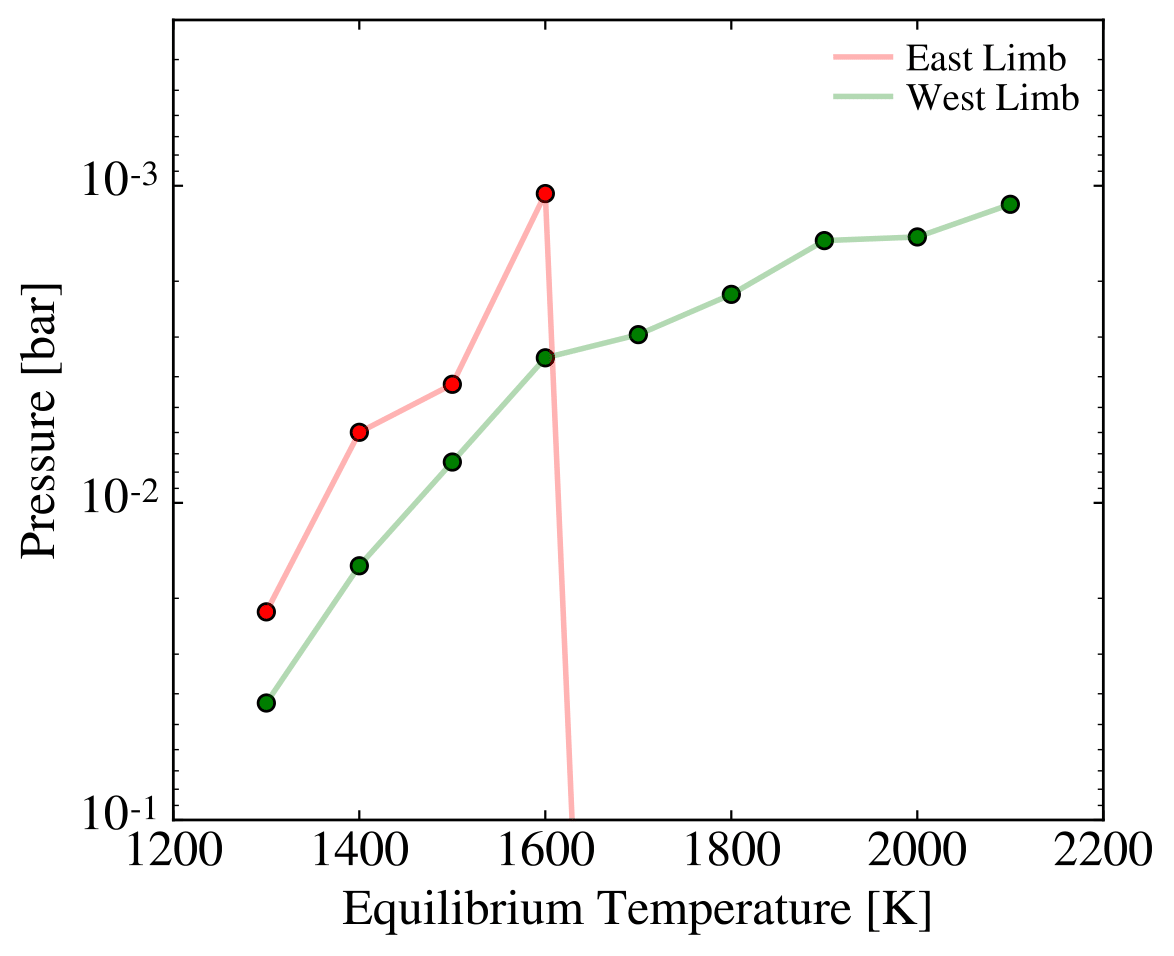}
\caption{The opaque cloud level at 3 $\mu$m for the east and west limbs as a function of equilibrium temperature. For T$_\text{eq} \le 1700$ K, the opaque cloud level at the east limb is higher in the atmosphere than at the west limb, despite there being a lower total cloud mass. }
\end{figure}\label{o_ew}

To derive the cloud particles' opacity we use complex refractive indices for MgSiO$_3$ from \citet{1975AJ.....80..587E} and \citet{1995A&A...300..503D}. For TiO$_2$ we use complex refractive indices from \citet{Kang1, Kang2}. Data for both clouds were compiled by \citet{2015A&A...573A.122W}. Our MgSiO$_3$ cloud particles are not homogenous since they have a core (TiO$_2$) and mantle (MgSiO$_3$) of different compositions. It is possible that these mixed cloud particles have different optical properties than those of pure MgSiO$_3$, however, any adjustments to their optical properties requires detailed modeling and/or laboratory experiments outside the scope of this work. Generally, as the size of the TiO$_2$ seed ($\sim 10^{-1}$ $\mu$m) is much smaller than the mantle of MgSiO$_3$ ($\sim 10$ $\mu$m), the optical properties should be similar to those of a pure MgSiO$_3$ particle. We therefore assume that the optical properties of the MgSiO$_3$ cloud particles with TiO$_2$ cores are roughly equivalent to those of pure MgSiO$_3$.

Given a wavelength and a complex refractive index, we can determine the extinction cross section ($\sigma_{ext}$) which in turn allows us to calculate the optical depth, ($\tau$). To compute $\sigma_{ext}$ we use \textit{bhmie}, a routine that uses Bohren-Huffman Mie scattering for a homogenous isotropic sphere to calculate scattering and absorption \citep{1983uaz..rept.....B}. This routine directly calculates the efficiency factor for extinction, efficiency factor for scattering, and the efficiency for backscattering. We use the extinction efficiency ($Q_{ext}$) to calculate $\sigma_{ext}$ via Equation \ref{extinct} where $a$ is the grain radius. 

\begin{equation}\label{extinct}
Q_{ext} = \frac{\sigma_{ext}}{\pi a^2}
\end{equation}

Given $\sigma_{ext}$ we calculate the optical depth for each particle size bin:
\begin{equation}\label{tau}
d\tau = n(l,r) \sigma_{ext}(r) dl,
\end{equation}

\noindent where $n(l,r)$ is the number density of cloud particles as a function of the path length of light ($l$) and particle radius ($r$). 

We then either take a cumulative sum of all of the vertical levels to find the emission optical depth (Nadir view) or we calculate the optical depth assuming transmission geometry along the line of sight. In the following sections we present the combined opacities of both the pure TiO$_2$ clouds and the MgSiO$_3$ clouds.

\subsection{Transmission Opacity}
We calculate the cloud particle contribution to the total atmospheric opacity in transmission at both the east and west limbs for each planet in our grid. The full transmission cloud opacities are shown in Figure \ref{trans_he}, where the white dotted line indicates the point in the atmosphere where the clouds become opaque ($\tau = 1$), which we refer to as the ``opaque cloud level". 

Of particular interest are observed differences between the two limbs as patchy cloud coverage has been shown to distinctly impact planetary transmission spectra \citep{2016ApJ...820...78L}. The clouds are optically thick at nearly all wavelengths for every equilibrium temperature at the west limb. The east limb, however, shows a clear progression from optically thick at lower equilibrium temperatures to optically thin at all wavelengths for equilibrium temperatures greater than 1800 K. 

While the east limb has less total cloud mass than the west limb, the opaque cloud level is located higher in the atmosphere for T$_\text{eq} \le 1700$ K. This trend is shown in Figure \ref{o_ew}. The east limb, therefore, appears more cloudy with increasing equilibrium temperature until the planet becomes too hot for clouds to form (T$_\text{eq} > 1700$ K). This is because the cloud base is higher in the atmosphere at locations with hotter temperature profiles. Therefore, if enough clouds can form such that the clouds become opaque they do so at higher levels, causing the cloud top to be located higher in the atmosphere.

\begin{figure}[tbp]
\epsscale{1.}
\plotone{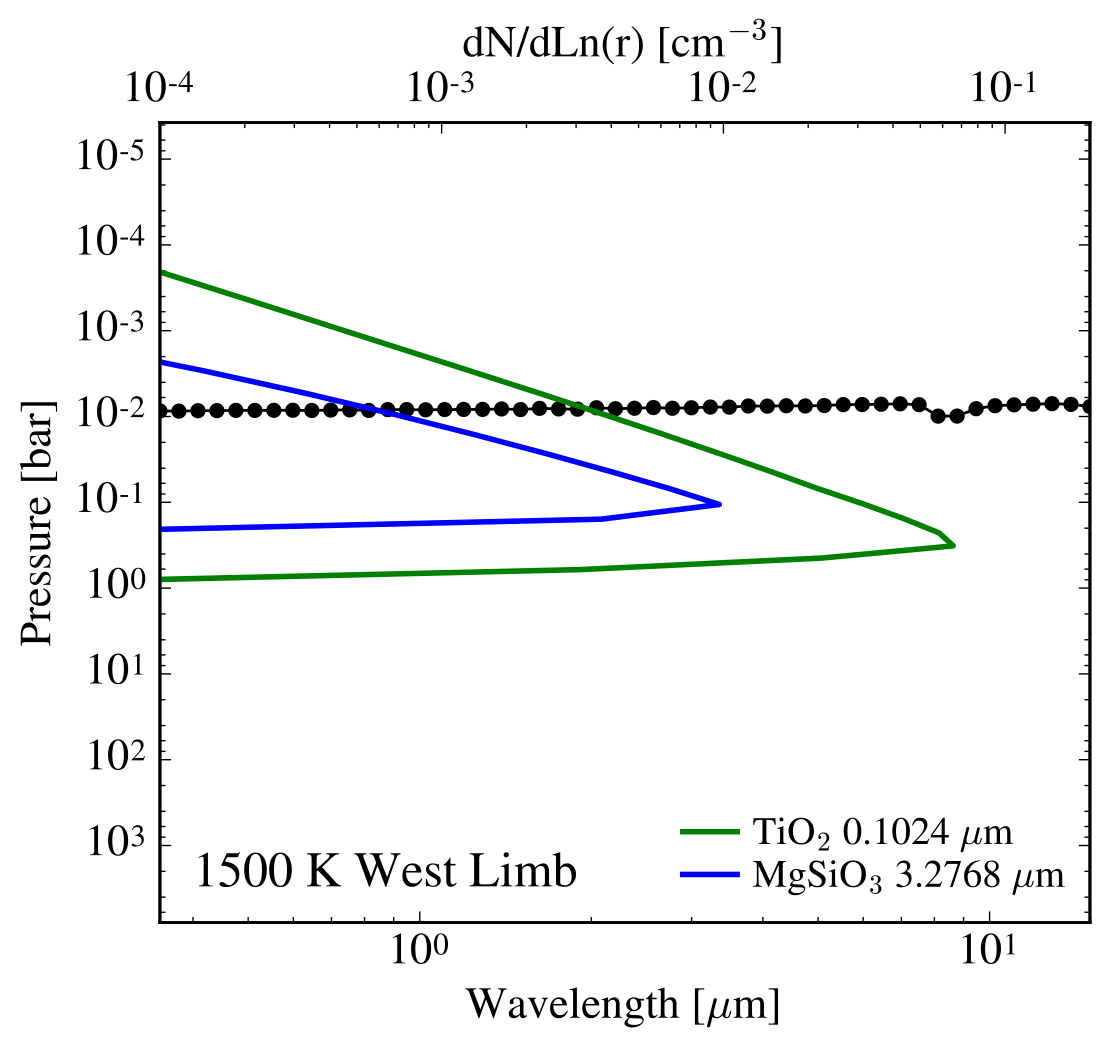}
\caption{The opaque cloud level across the full wavelength range (black, dotted) as compared to the total distribution of cloud particles in terms of number density ($dN/dLn(r)$) for the particle size bins (see legend) that contribute the most to the cloud opacity. Shown is the case of a 1500 K hot Jupiter at the west limb. The distribution of titanium clouds is shown in green and the distribution of silicate clouds is shown in blue. There is not an increase in particle density near the cloud top.}
\end{figure}\label{transmass}

\begin{figure}[tbp]
\epsscale{1.2}
\plotone{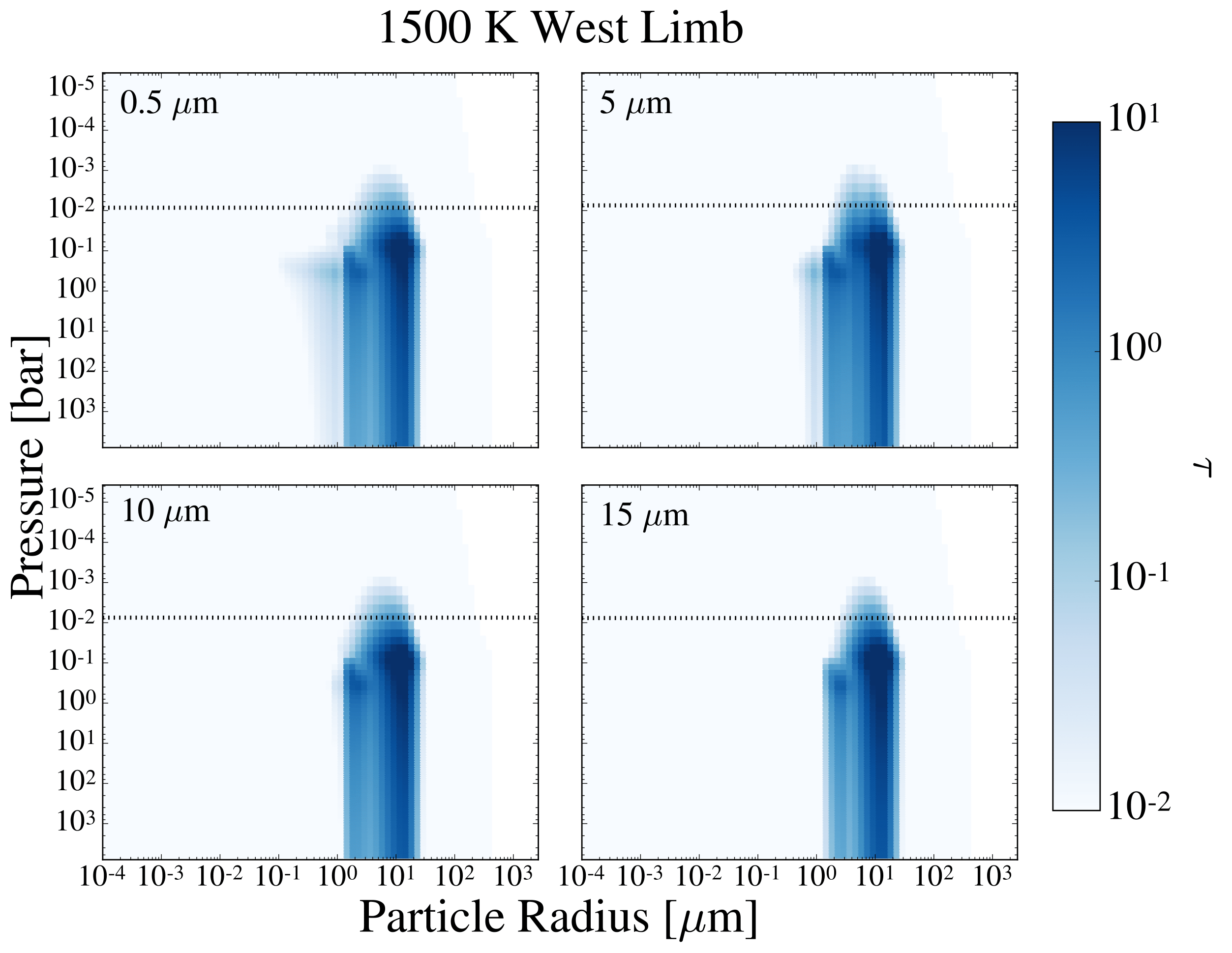}
\caption{The contribution to the total cloud transmission opacity from each cloud particle size bin as a function of atmospheric pressure for 4 representative wavelengths. Shown is the case of a 1500 K hot Jupiter at the west limb. The black dashed line indicates the opaque cloud level at a given wavelength. The large cloud particle sizes cause the cloud opacities to be flat across a broad wavelength range.}
\end{figure}\label{flatspec}
     
\begin{figure}[tbp]
\epsscale{1.}
\plotone{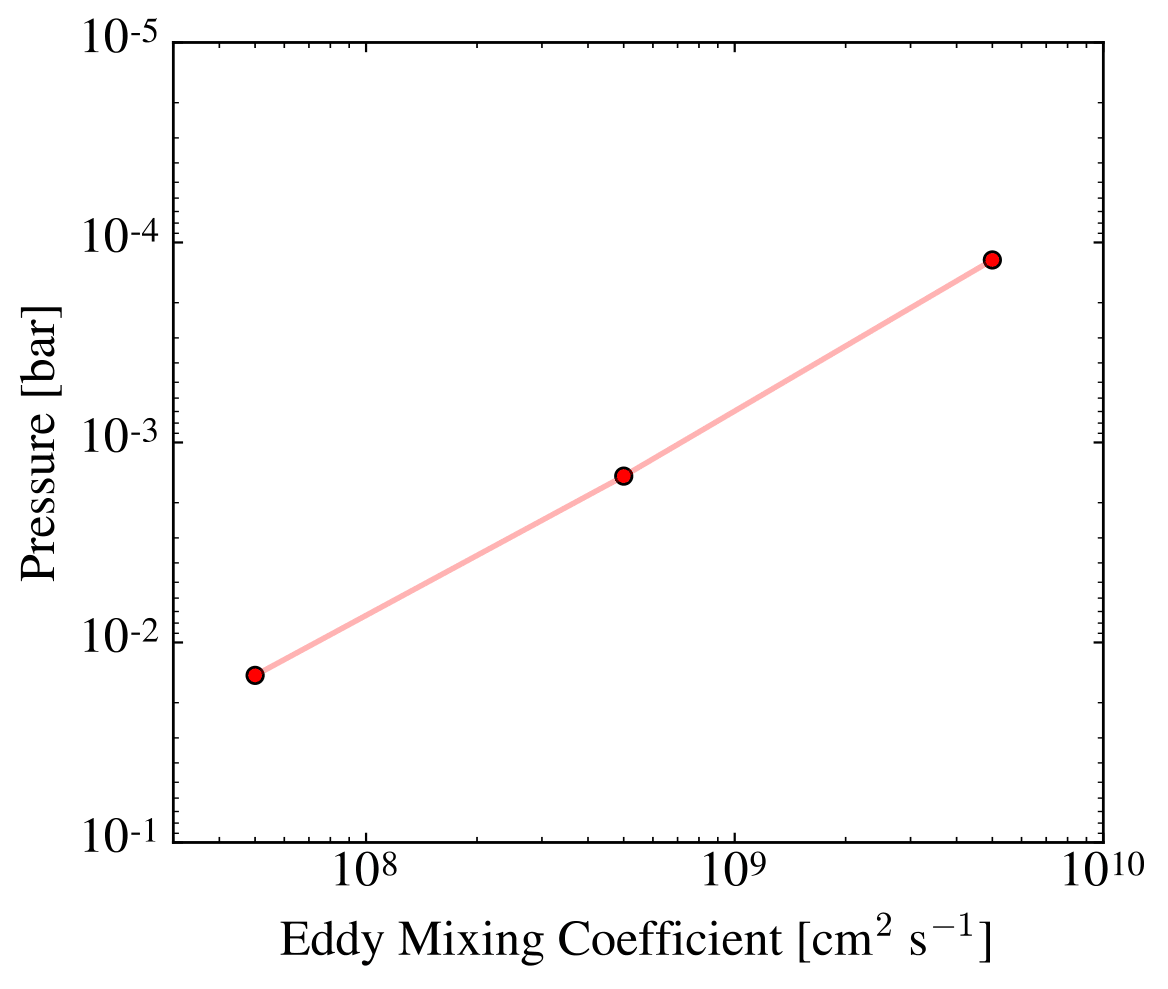}
\caption{The opaque cloud level at 3 $\mu$m for a 1700 K hot Jupiter as a function of atmospheric vertical mixing. Increasing the vertical mixing coefficient by an order of magnitude correspondingly raises the location of the opaque cloud level by roughly an order of magnitude in pressure.}
\end{figure}\label{kzz_clo}

\begin{figure*}[tbp]
\epsscale{1.2}
\plotone{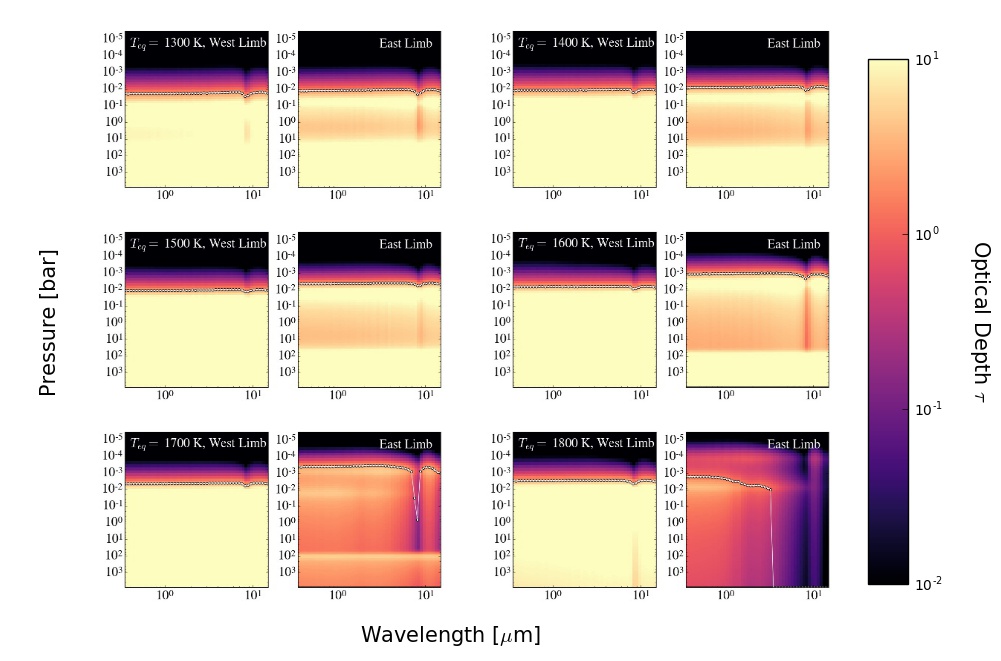}
\caption{Cloud transmission opacities for the case of the low entropy interior. The white dotted line represents the opaque cloud level. There are noticeable hemispheric differences between the east and west limb for planets with equilibrium temperatures of 1700 and 1800 K.}
\end{figure*}\label{trans_le}

\begin{figure*}[tbp]
\epsscale{1.2}
\plotone{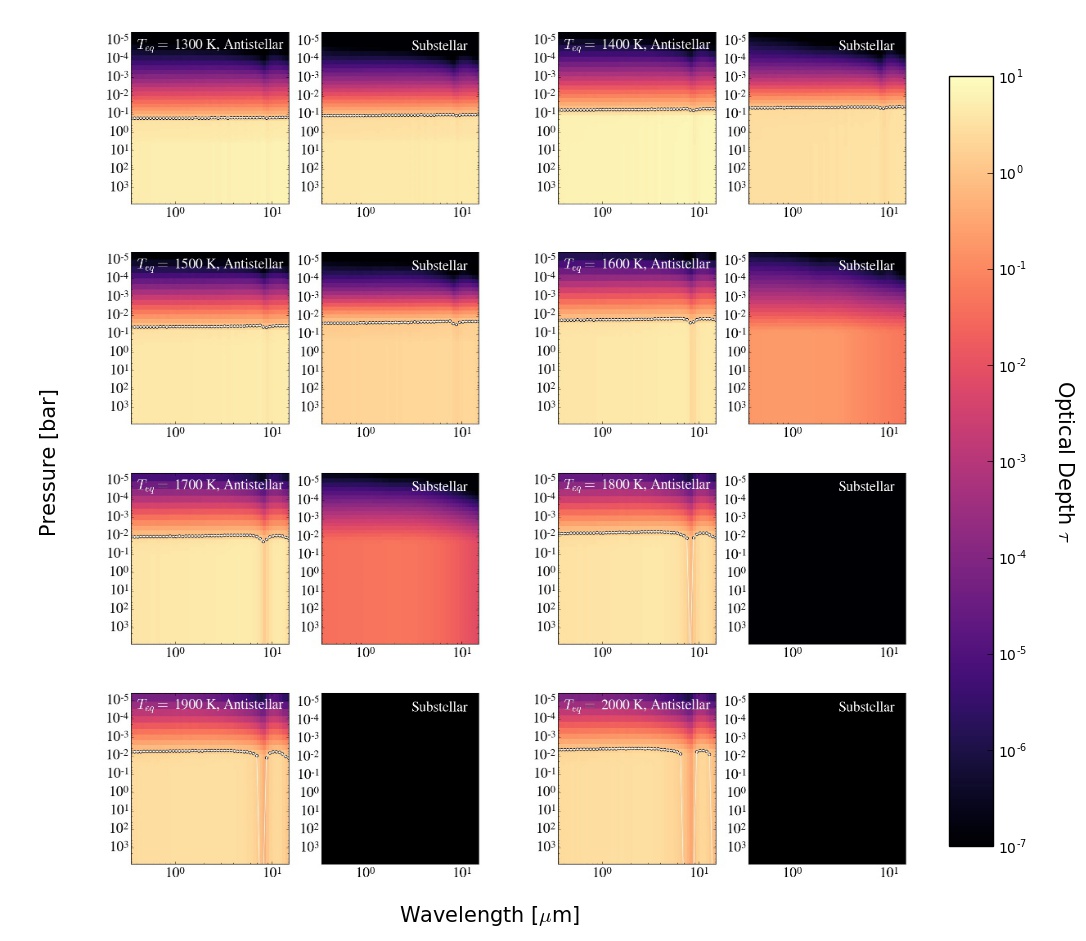}
\caption{Cloud nadir view (emission) opacities for the case of the high entropy interior. The white dotted line represents the opaque cloud level. The clouds are optically thick along the antistellar point and optically thin for planets with temperatures greater than 1500 K at the substellar point.}
\end{figure*}\label{nadir_he}

\begin{figure*}[tbp]
\epsscale{1.2}
\plotone{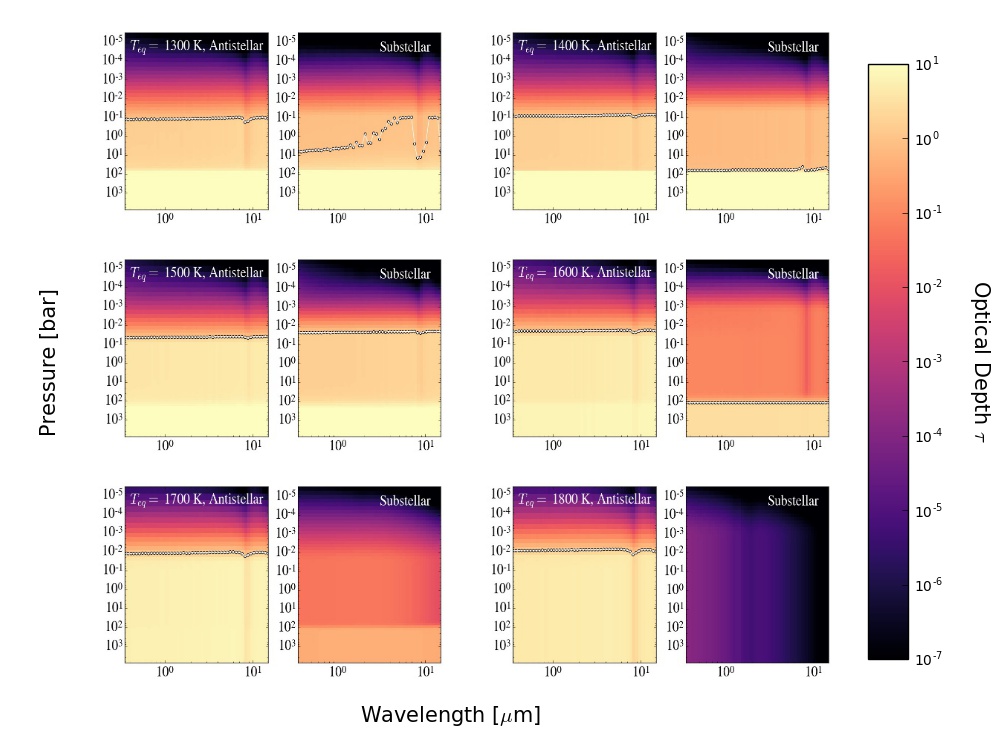}
\caption{Cloud nadir view (emission) opacities for the case of the low entropy interior. The white dotted line represents the opaque cloud level. The clouds are optically thick in the deep atmosphere for equilibrium temperatures less than 1700 K due to the presence of a deep cold trap. These emission opacities significantly differ from the high entropy interior case.}
\end{figure*}\label{nadir_le}

\begin{figure}[tbp]
\epsscale{1.}
\plotone{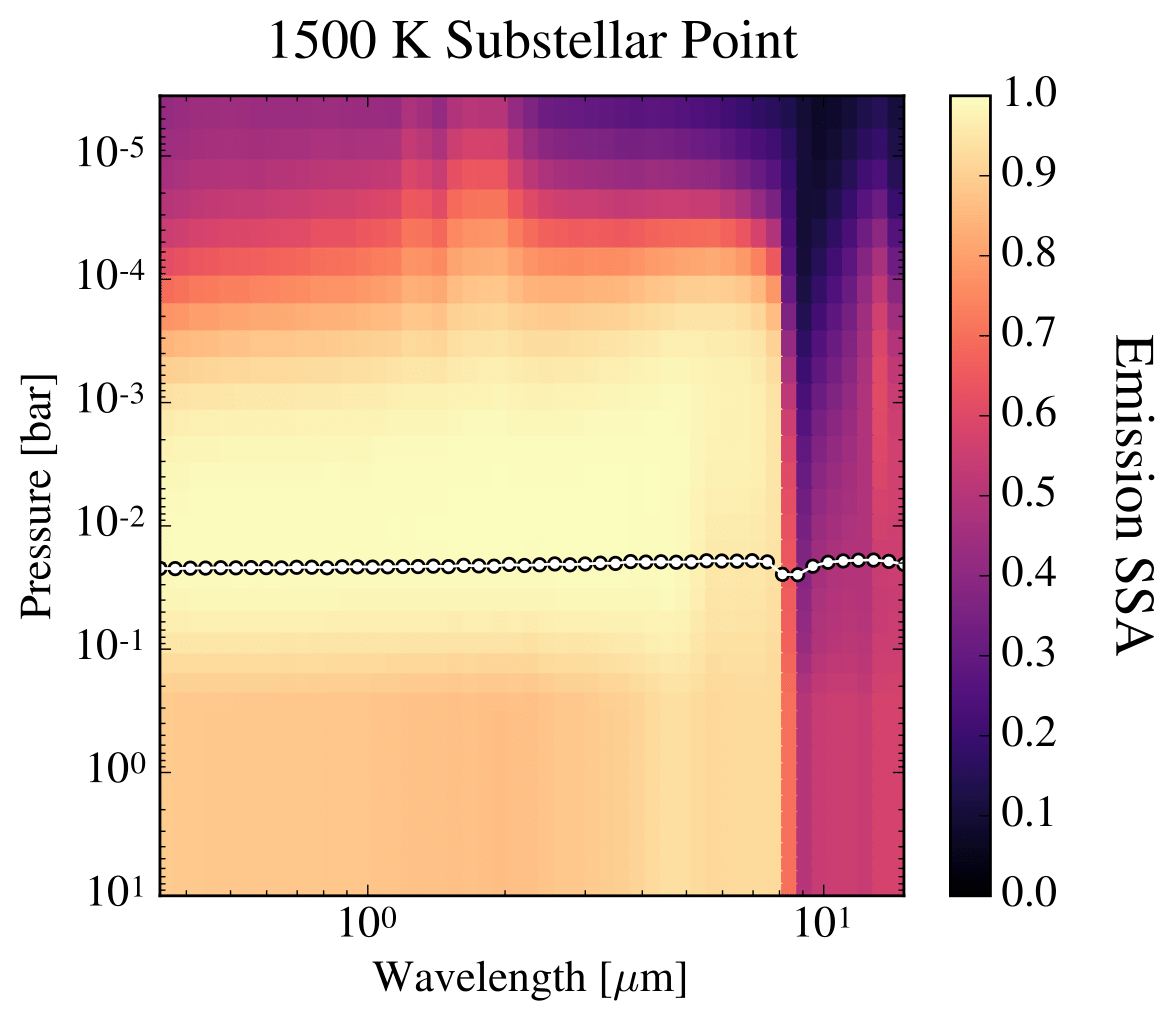}
\plotone{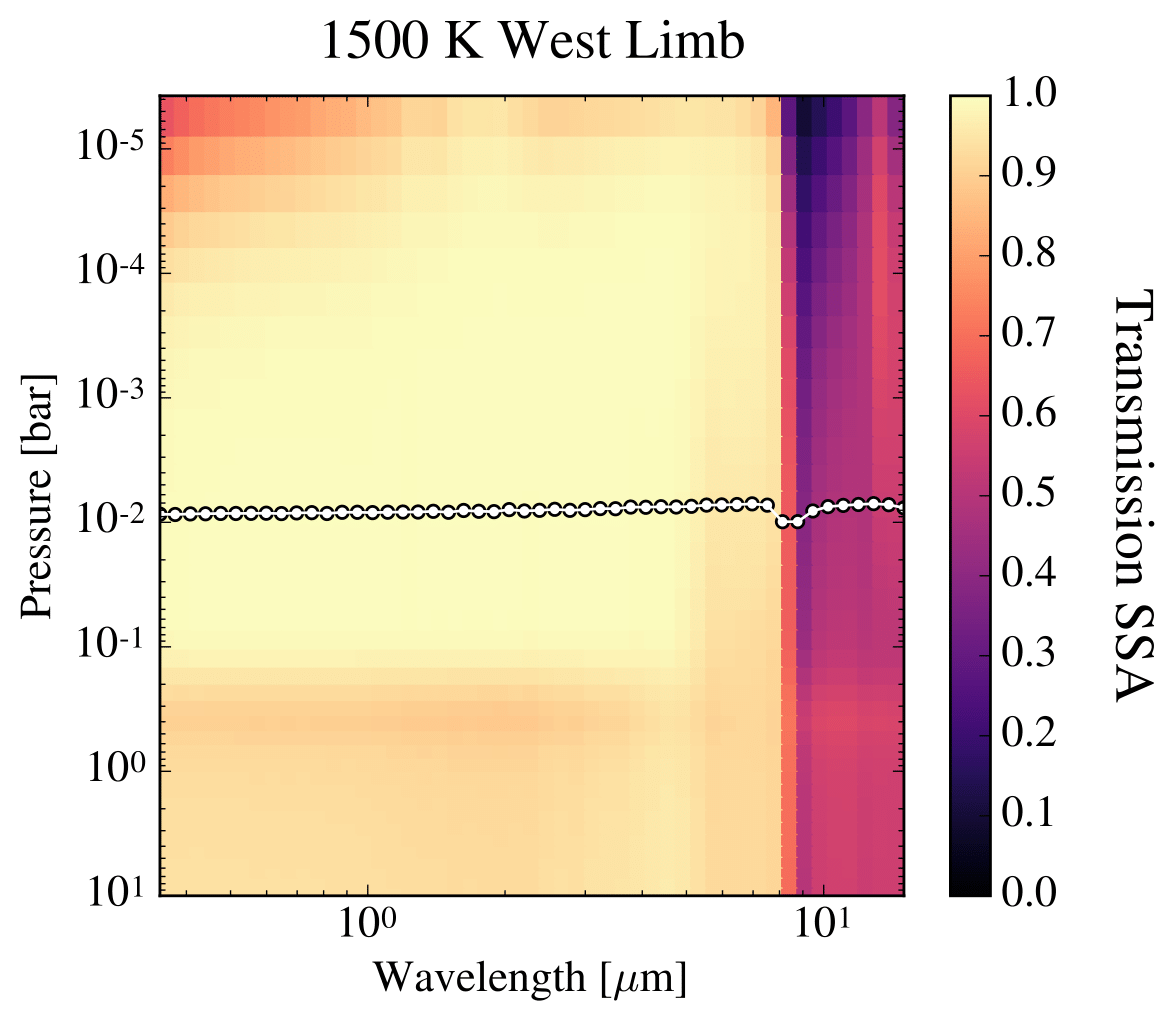}
\epsscale{1.05}
\plotone{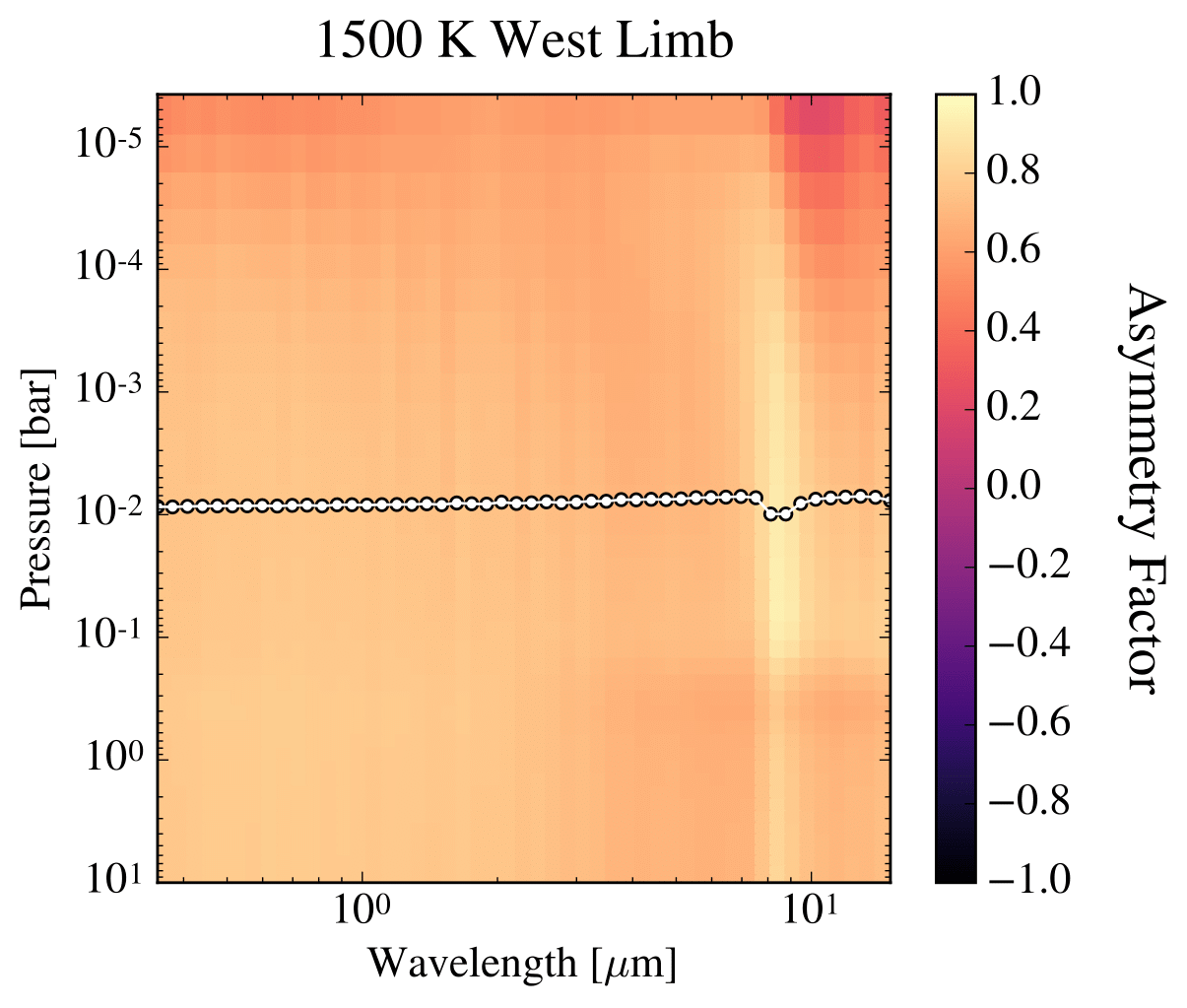}
\caption{Scattering properties of titanium and silicate cloud particles in a 1500 K planet with a high entropy interior. The white dotted lines indicate the opaque cloud level at each wavelength. Across all sampled wavelengths and pressures, titanium and silicate clouds are strong forward scatterers. This is particularly true for wavelengths shorter than 10 $\mu$m. Top Panel: Cumulative single scattering albedo as a function of wavelength and pressure as observed in emission. Middle Panel: Single scattering albedo as observed in transmission. Bottom Panel: Asymmetry parameter as observed in transmission.}
\end{figure}\label{ssa_fig}

\begin{figure*}[tbp]
\epsscale{.75}
\plottwo{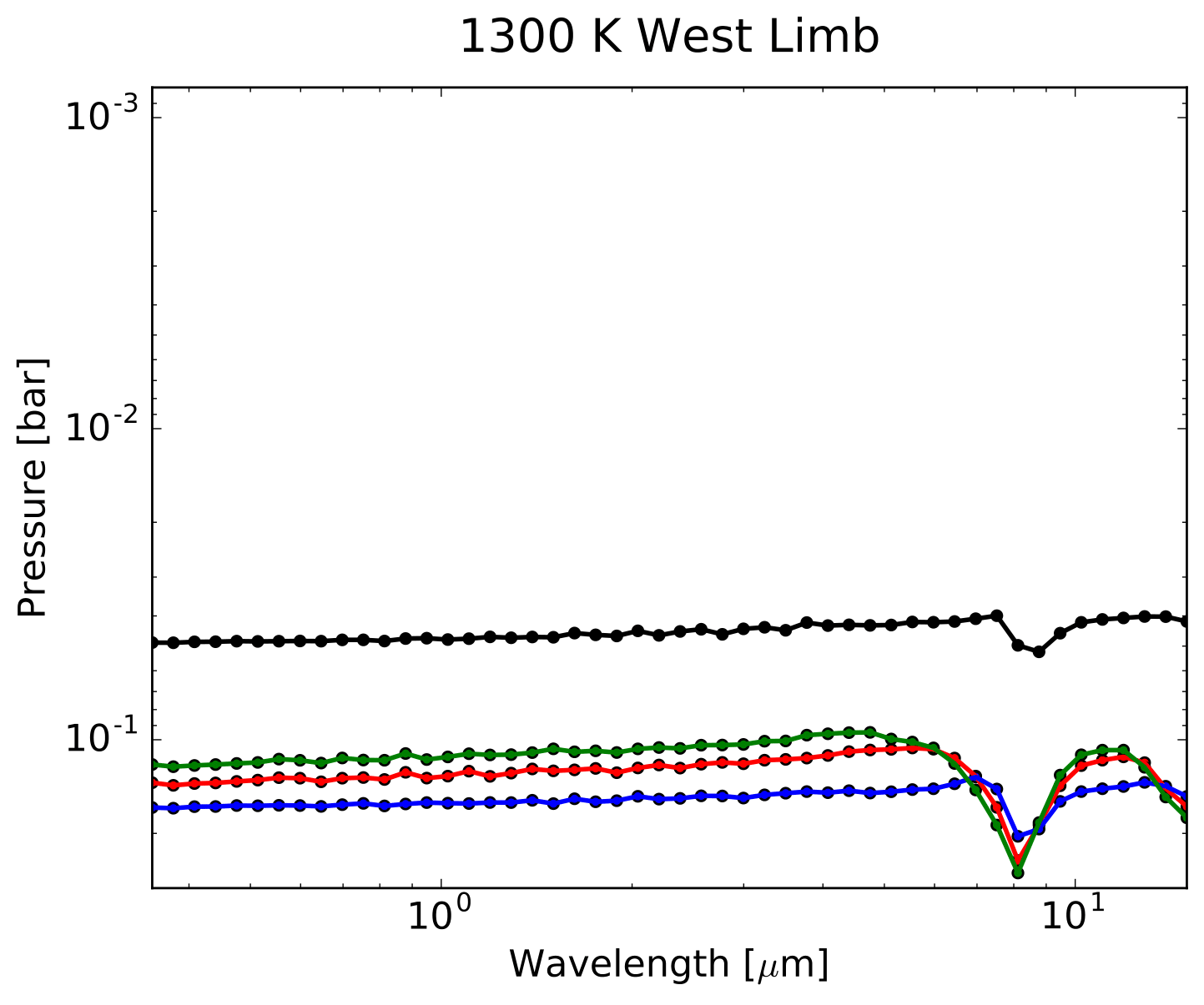}{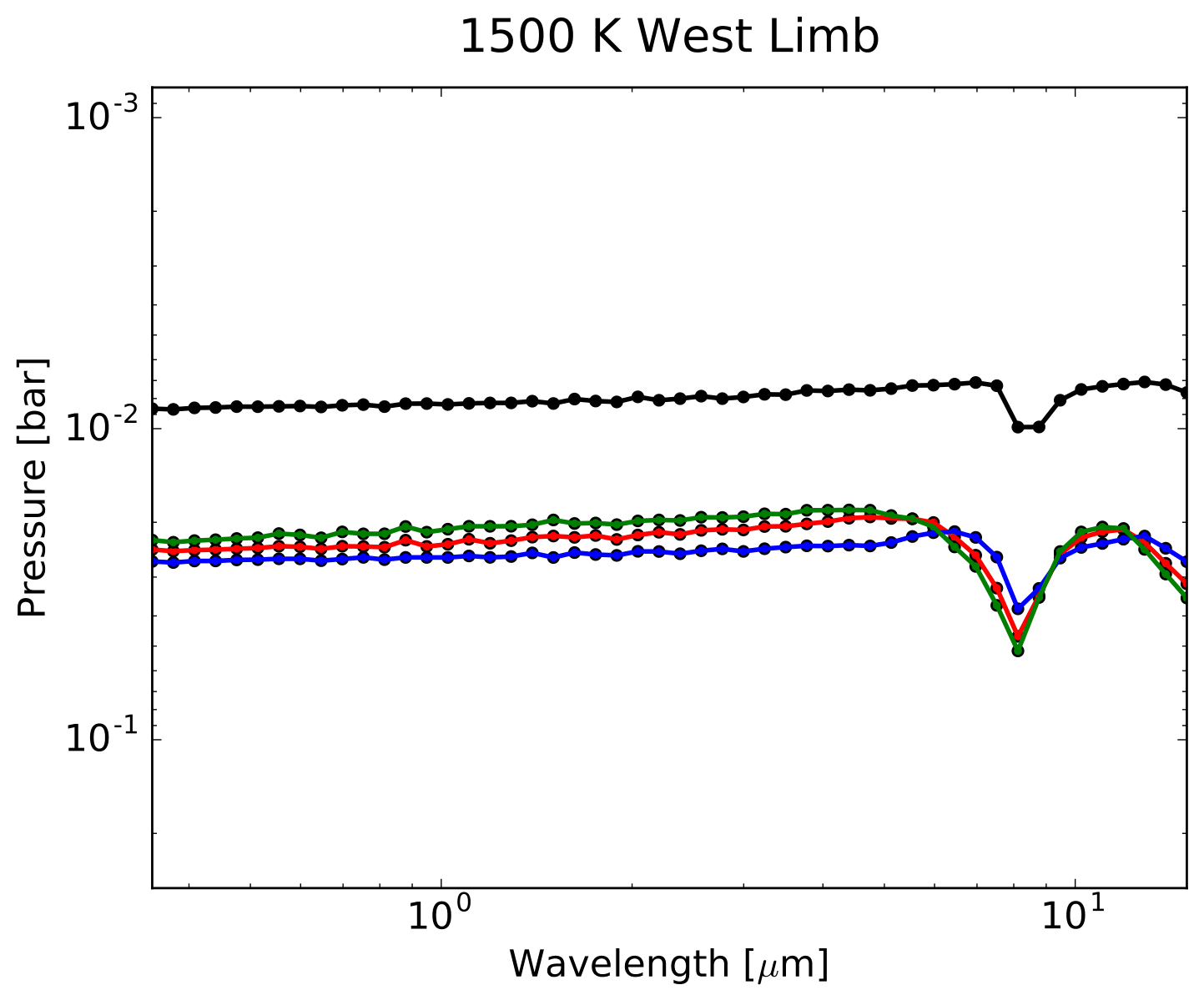}
\plottwo{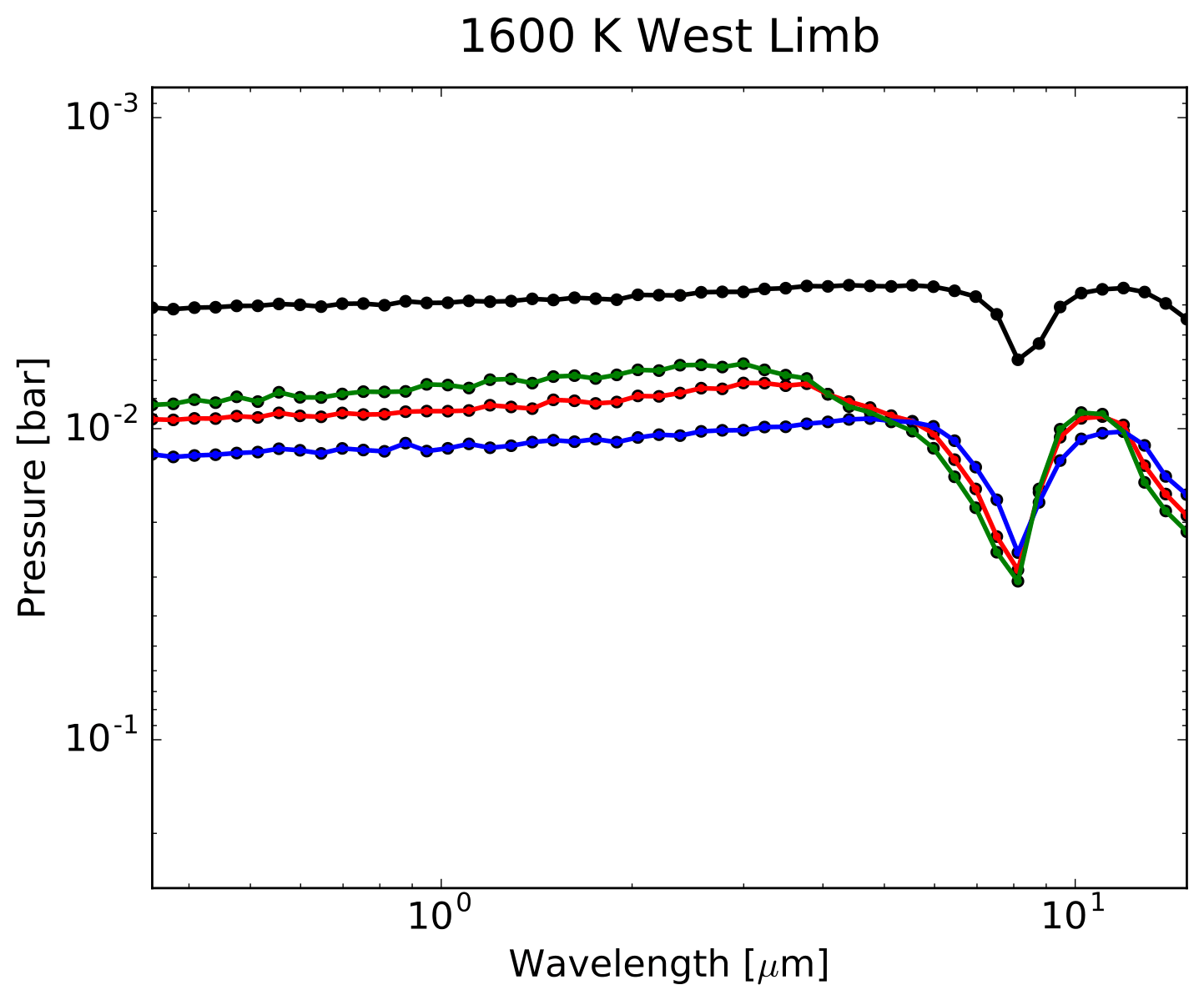}{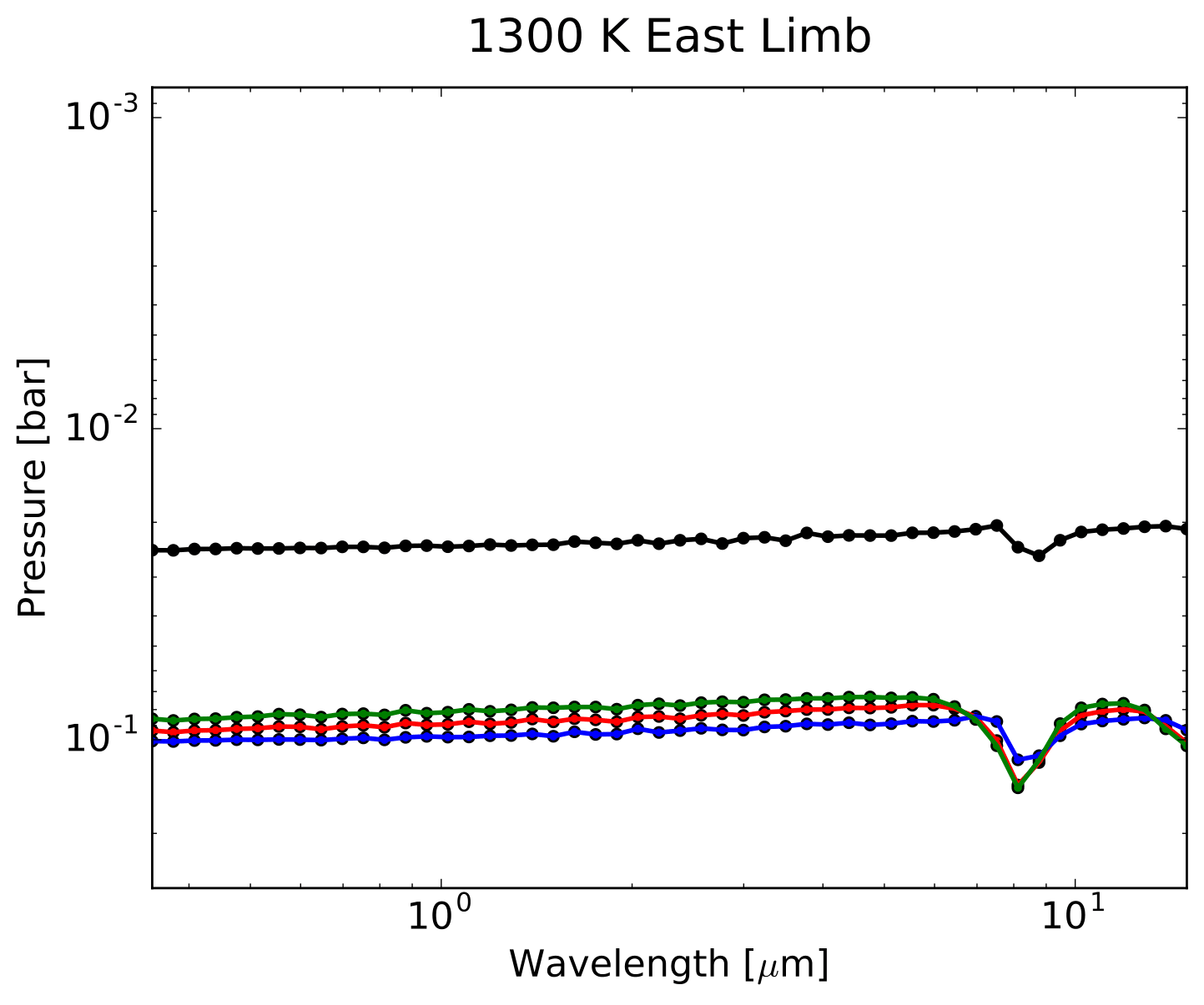}
\plottwo{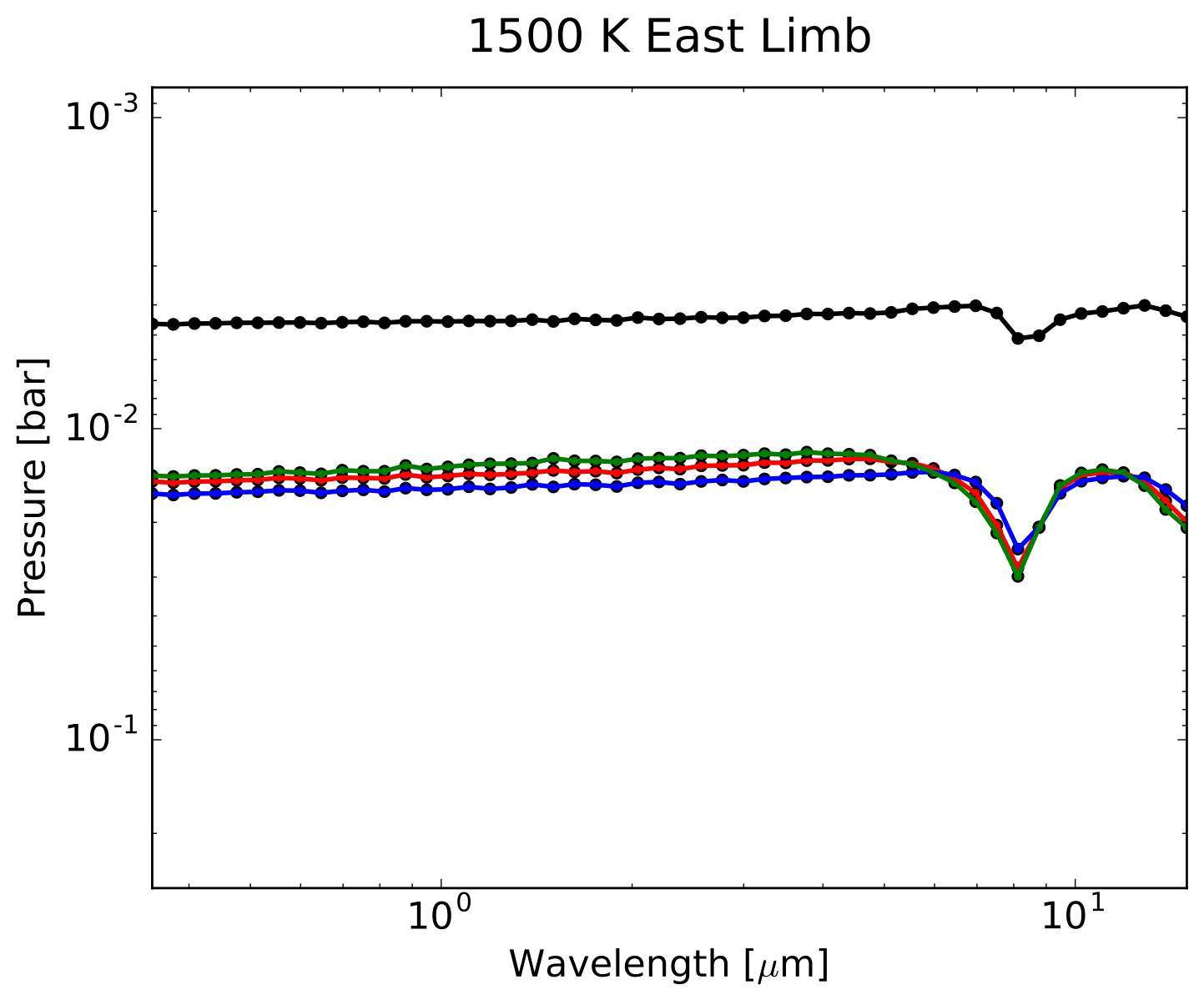}{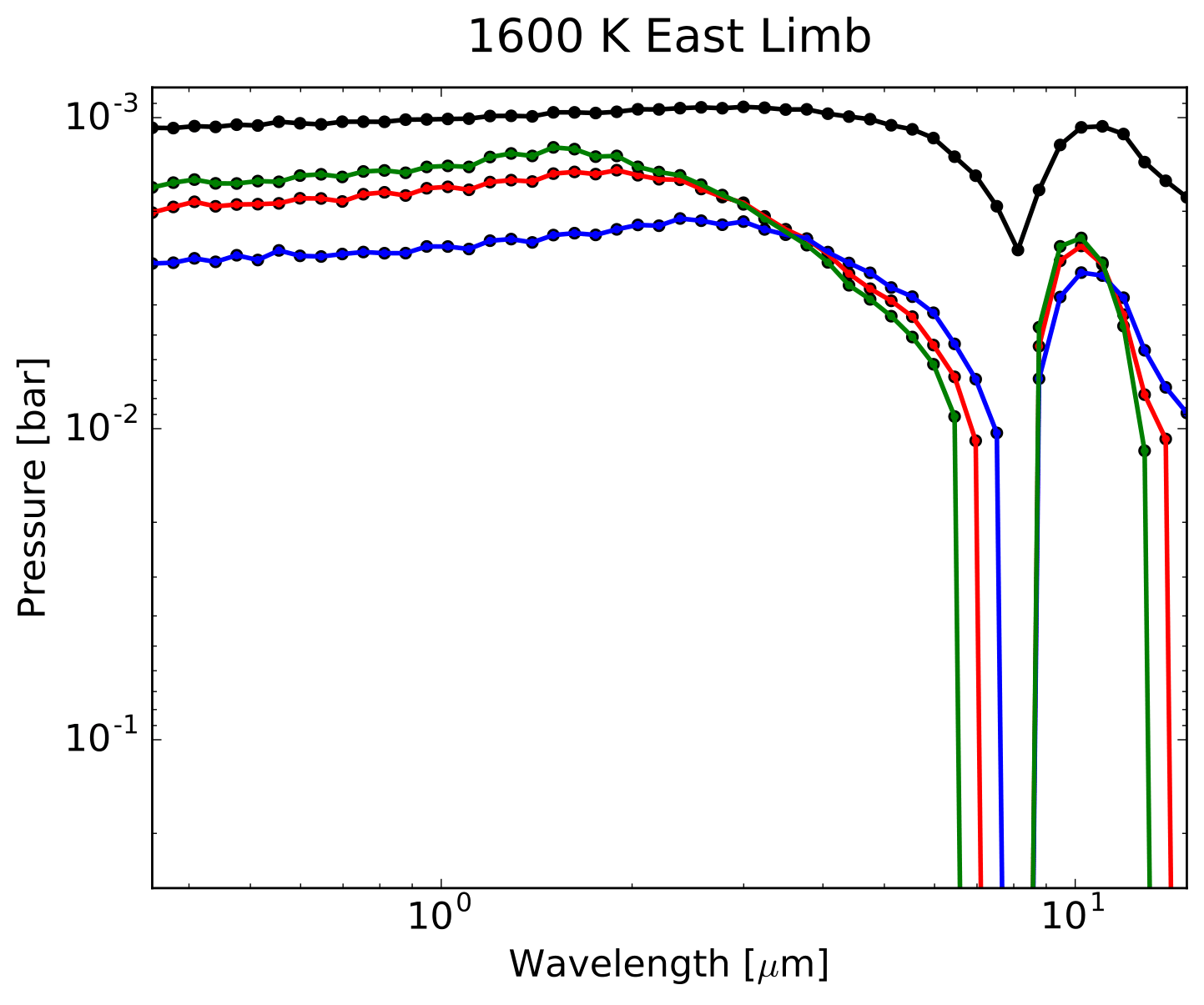}
\caption{The opaque cloud level as a function of wavelength and pressure calculated using the full particle size distribution (black), a representative mass weighted mean particle size (red), a cross section weighted mean particle size (green), and an area weighted mean particle size (blue). Methods that use a mean particle size typically underestimate the cloud opacity by a factor of $\sim 3$ to $5$ or more. }
\end{figure*}\label{mtrans}

For all planets at the west limb and for planets with T$_\text{eq} \le 1700$ K at the east limb, the cloud opacities are characteristically flat and featureless across a large wavelength range. One exception to this is a silicate absorption feature at 10 $\mu$m and a relatively clear region of the atmosphere at $\sim 8$ -- $9 \;\mu$m. These features in the infrared mirror the features in the refractive index of MgSiO$_3$ \citep[see][]{2015A&A...573A.122W}. Another exception is the east limb of the 1800 K planet, where only smaller TiO$_2$ cloud particles are abundant. The opacity profile in this case is reminiscent of the observed slope in transmission spectra at short wavelengths \citep[e.g.,][]{2016Natur.529...59S, 2017MNRAS.468.3907K}. 

Our calculated cloud opacities are gray across a large wavelength range due to the presence of relatively large cloud particles. Clouds can appear gray either due to having large sized particles or to a sharp increase in the number density of small particles near the cloud top \citep{2015arXiv150407655B}. As shown in Figure \ref{transmass}, where we plot the cloud particle number density for the size bin that contributes the most to the opacity above the opaque cloud level, there is no such increase in cloud particles near the cloud top. The cloud particles instead appear opaque due to their large size, as indicated by Figure \ref{flatspec}, where we plot each size bin's contribution to the total opacity for 4 representative wavelengths. While the opacity of each particle size bin varies with wavelength, the presence of relatively large particles causes the clouds to be gray across our full wavelength range. 

We therefore conclude that it is unlikely that MgSiO$_3$ or TiO$_2$ clouds are responsible for the observed Rayleigh scattering slope in the optical confirming the result found using a different framework in \citet{2017A&A...601A..22L}. For MgSiO$_3$ this is due to the inefficient rate of homogenous nucleation at small sizes as well as this species's efficient growth. These two effects skew the particle distribution towards larger radii. While TiO$_2$ does nucleate homogeneously at small sizes, the number density of small cloud particles in the upper atmosphere is insufficient to produce the observed Rayleigh slope. The Rayleigh-like slope requires the presence of many small cloud particles in the upper atmosphere which may still occur for these cloud species if there is an enhanced presence of CCN (such as photochemical hazes) such that the gas supply is preferentially used for nucleation and growth is starved.

MgSiO$_3$ cloud particles contribute to silicate dust features at $\sim 10\;\mu$m that may be observable with JWST. However, these features are not as large as predicted by previous work \citep[e.g.,][]{2015A&A...573A.122W} due to the presence of large cloud particles. Furthermore, the cloud particles in our modeling are sufficiently opaque that we do not expect that signatures of a cloud base will be observable, as proposed by \citet{2014ApJ...789L..11V}. The possible exception to this may be for clouds along the east limb for T$_\text{eq} > 1800 K$ (see Figure \ref{trans_he}), however, this would depend on the magnitude of the gas opacity which we do not take into account.

The strength of vertical mixing in an atmosphere will determine the location of the opaque cloud level. This is shown in Figure \ref{kzz_clo}, where we plot the opaque cloud level pressure at 3 $\mu$m as a function of vertical mixing for a 1700 K hot Jupiter. Increasing the vertical mixing coefficient by an order of magnitude correspondingly raises the opaque cloud level by roughly an order of magnitude in pressure.

We now examine the cloud transmission opacities for the low entropy interior cases to understand the efficiency of the deep cold trap. The deep cold trap is inefficient at most locations and equilibrium temperatures at limiting cloud formation in the upper atmosphere, such that the upper level clouds are optically thin in transmission. These opacities are shown in Figure \ref{trans_le}. This is particularly true for the west limb, where the cloud particles high in the atmosphere are nearly as opaque as the cloud particles in the deep atmosphere. The cold trap is more efficient along the east limb. However, this effect is not typically large enough to significantly impact the location of the opaque cloud level as compared to the case of the high entropy interior. 

The presence of a deep cold trap will likely be of increased importance in atmospheres with inefficient vertical mixing. This is because gas will be comparatively slow to diffuse to the upper atmosphere and replenish the supply of condensible material. Limiting the supply of cloud forming material in the upper atmosphere thus strengthens the effect of the deep cold trap. Furthermore, for planets with temperature profiles similar to those of the low entropy interior case, the presence of two cloud decks could complicate observational determinations of total cloud mass or atmospheric metallicity, as the deep clouds do not contribute to the observed opacity.

\subsection{Nadir View Opacity}\label{nadir}
We calculate the cumulative optical depth of the clouds in a nadir viewing geometry for the antistellar and substellar points. This geometry is equivalent to a planet viewed in emission. 

All planets in our high entropy grid are opaque in emission at the antistellar point with an opaque cloud level that ranges from $10^{-1}$ -- $10^{-2}$ bar as shown in Figure \ref{nadir_he}. Planets with equilibrium temperatures greater than 1500 K are clear at the substellar point across all wavelengths. For planets with temperatures less than 1500 K, the opaque cloud level at the substellar point is at roughly the same location as it is at the antistellar point.

The opacity profile in emission for these clouds is again flat and featureless across a broad wavelength range with the exception of a 10 $\mu$m absorption feature for planets with equilibrium temperatures greater than 1600 K. This absorption feature is accompanied by a narrow wavelength range for which the clouds are relatively clear, from roughly $8$ -- $9\;\mu$m, again mirroring features in the refractive index for MgSiO$_3$. 

There is a difference between the high and low entropy cases in emission, as shown in Figure \ref{nadir_le}. The deep cold trap causes the opaque cloud level to be located lower in the atmosphere for the low entropy interior at the substellar point for equilibrium temperatures less than 1700 K. For these planets, the clouds in the upper atmosphere are clear across a broad wavelength range. There are also distinctive infrared features at the antistellar point in the 1300 and 1400 K planets and at the substellar point in the 1500 K planet that are not present in the case of the high entropy interior. 

This difference in emission opacities demonstrates that observable cloud properties can be an indicator of the internal thermal structure of a planet and can even distinguish between different planetary inflation mechanisms. We therefore predict that differences in the internal structure of a hot Jupiter should be most readily observable in emission, particularly as this viewing geometry is a more sensitive probe of cloud mass \citep{2005MNRAS.364..649F}.

Interestingly, the nadir view cloud opacity at the substellar point for a low entropy 1500 K planet are opaque slightly higher in the atmosphere than at the antistellar point. This location and equilibrium temperature represents a special case in which the atmosphere is hot enough such that only titanium clouds will form while the upper atmosphere is cool enough such that both silicate and titanium clouds are able to form (see Figure \ref{he_prof}). This means that a supply of SiO gas is able to reach the upper atmosphere and form enough large clouds such that the upper cloud deck becomes opaque. This, along with the 1600 K case, are the only substellar cases in our modeling where both cloud species form while only one species is cold trapped. This differs from the substellar point of the 1400 K planet where both cloud species are cold trapped and both are also able to form in the upper atmosphere. The lower cold trap in this case limits cloud formation in the upper atmosphere such that the population of high clouds is optically thin in a nadir viewing geometry. The 1500 K planet also differs from the substellar point of the 1600 K planet where the supersaturation of silicate clouds in the upper atmosphere is significantly lower, resulting in the formation of only optically thin clouds high in the atmosphere. For the 1700 K planet at the substellar point only titanium clouds are cold trapped and only an optically thin layer of titanium clouds form in the upper atmosphere.

\subsection{Single Scattering Albedo}\label{SSA}
Here we determine whether scattering plays an important radiative role for titanium and silicate clouds. We do this through calculating the single scattering albedo (SSA) of our cloud particle size distributions for both the nadir and transmission viewing geometry. The SSA is the ratio of the scattering efficiency to the total extinction efficiency. When the SSA is close to unity the particles are strong scatterers and when the SSA is close to zero the particles are strong absorbers. This is particularly important as previous work by \citet{2012MNRAS.420...20H} has shown that a consideration of scattering effects from clouds and hazes will modify the inferred temperature profile of a planet. 

For all wavelengths and for all cases with appreciable clouds, scattering plays an important role in emission at wavelengths shorter than 10 $\mu$m. This is shown in the top panel of Figure \ref{ssa_fig} for the nadir viewing geometry for a 1500 K hot Jupiter at the antistellar point. While scattering is dominant at short wavelengths, it continues to play a significant role across the full wavelength range considered. 

We also derive the single scattering albedo for our cloud particle distributions as viewed in transmission. While scattering effects are not typically calculated in modeling transmission spectroscopy, previous work has shown that scattering in transmission may be important in understanding spectra \citep{2017ApJ...836..236R}. 

For all of our planet cases, scattering in transmission is significant. For example, the SSA for the representative case of a 1500 K hot Jupiter at the west limb for the high entropy interior case ranges from $\sim$ 1 at shorter wavelengths to $\sim$ 0.5 at wavelengths larger than 10 $\mu$m, as shown in the middle panel of Figure \ref{ssa_fig}. This indicates that scattering is important in transmission calculations for silicate and titanium clouds. 

This is further confirmed through a calculation of the asymmetry parameter, as shown in the bottom panel of Figure \ref{ssa_fig}. The asymmetry factor indicates a particle's tendency to forward scatter, where particles with an asymmetry parameter of unity are strongly forward scattering. Across all wavelengths and relevant pressures, titanium and silicate clouds are strong forward scatterers. This again indicates the importance of considering scattering effects in relevant transmission calculations.

\subsection{The Impact of Using Realistic Particle Size Distributions}
Using the fully resolved cloud particle size distribution has a distinct impact on derived atmospheric observables, indicating that detailed cloud modeling is essential for understanding the atmospheres of hot Jupiters. We confirm this by calculating the amount by which the full particle size distribution changes the opaque cloud level in transmission as compared to a calculation using the mean particle size alone.

For this comparison, we calculate three different mean particle sizes for each cloud species at each vertical level in the atmosphere. We calculate the mass weighted mean particle size, the area weighted mean particle size, and the cross section weighted mean particle size. We assume for each mean particle size that the total cloud mass is the same as for the full particle size distribution calculated using CARMA. We are then able to calculate transmission opacities for the resulting cloud particle distributions. A comparison of the opaque cloud level for these four methods is shown in Figure \ref{mtrans}. 

All methods that use a mean particle size underestimate the cloud opacity by a factor of $\sim 3$ -- $5$ or more. The reason for this is that all methods of deriving a mean particle size tend to skew towards a large mean value that neglects the substantial contributions to the opacity from smaller particles in the size distribution.

At higher equilibrium temperatures, the cross section weighted mean particle size nearly matches the opaque cloud level derived using the full size distribution at short wavelengths. For lower equilibrium temperatures, all three mean particle size methods underestimate the opacity by roughly the same amount across all wavelengths. This shows that a consideration of the full cloud particle size distribution is essential for accurate spectral analysis. 

A consideration of the full particle size distribution gives further insight into the process by which cloud particles impact atmospheric observables. For instance, the particle size that contributes the most to the cloud opacity depends on the cloud particle size distribution. \citet{2015A&A...573A.122W} find that the largest particle size contributes the most to the opacity for a log-normal particle size distribution. However, we find that the largest particle size does not always contribute the most to the opacity for our fully resolved size distributions. This effect is shown in Figure \ref{flatspec}, where we examine each particle size bin's contribution to the total cloud opacity in transmission at four representative wavelengths. In this case, the largest particles do not contribute the most to the opacity at the opaque cloud level. 

In cases where there are significant populations of both large and small cloud particles, it is possible for large cloud particles to dominate the cloud opacity and effectively obscure cloud material in the deep atmosphere. This effect is wavelength dependent and can again be seen in Figure \ref{flatspec}. In cases such as these, careful modeling of observations is necessary to accurately determine the total cloud mass and/or cloud dependent metallicity.  

\subsection{Comparison to Observational Inferences}
We now provide a brief comparison of our more general results to several observed planets. The presence of a gray cloud deck is a necessary feature to understand the transmission spectra of most hot Jupiters. Our calculations confirm that the presence of a gray cloud deck should be ubiquitous across a range of planetary temperatures. 

We are able to reproduce the opaque cloud deck for WASP 43b \cite[T$_\text{eq}$ = 1440 K,][]{2014ApJ...781..116B} with a consistent location of the cloud top (opaque cloud level) of P $= 10^{-1^{+1.1}_{-0.8}}$ bar as given in \citet{2014ApJ...793L..27K}. Recent retrievals in transmission for WASP 17b (T$_\text{eq}$ = 1740 K) and WASP 19b (T$_\text{eq}$ = 2050 K) indicate the presence of a cloud top at roughly $10^{-3}$ bar \citep{2017ApJ...834...50B}, consistent with our derived cloud tops in transmission for similar equilibrium temperatures. Similarly, the presence of a gray cloud deck in the mid-atmosphere of HD 209458b (T$_\text{eq}$ = 1400 K) necessary to understand the transmission spectra \citep{2015arXiv150407655B} naturally arises from our calculations. Rough constraints on the cloud top of 200 mbar to 0.01 mbar from \citet{2015arXiv150407655B} are roughly consistent with the cloud top inferred from our models for a hot Jupiter with a similar equilibrium temperature. 

Additionally, WASP 2b (T$_\text{eq}$ = 1284 K), WASP 24b (T$_\text{eq}$ = 1583 K), and HAT-P 5b (T$_\text{eq}$ = 1713 K) have notably flat spectra across a broad wavelength range consistent with the presence of a gray cloud deck as derived in our calculations for planets of similar equilibrium temperatures \citep{2017MNRAS.472.3871T}. WASP 31b (T$_\text{eq}$ = 1580 K) also shows damped spectral features, again indicating the presence of a gray cloud deck \citep{2016Natur.529...59S}. 

Our derived titanium and silicate cloud populations do not produce the Rayleigh scattering slope at short wavelengths as observed in the transmission spectra of many hot Jupiters \citep{2016Natur.529...59S}. This confirms the result from \citep{2017A&A...601A..22L} where they are unable to fully reproduce observational slopes using condensational clouds. This slope might instead be due to the large abundance of small photochemical haze particles \citep{2017ApJ...847...32L} or the presence of a different cloud species. Or the Rayleigh slope could also be due in part to contaminating stellar activity \citep[e.g.,][]{2014ApJ...791...55M}.

\section{Summary and Conclusions}\label{sum}
We present the first bin-scheme microphysical model of cloud formation on hot Jupiters. This framework can predict detailed cloud properties from first principles. In particular, this approach enables a derivation of the fully resolved cloud particle size distribution that will become increasingly important as atmospheric datasets continue to improve. 

In this work we summarize the theory of cloud formation from the microphysical perspective, with a particular emphasis on the processes of nucleation, condensational growth, and evaporation. We then detail modifications made to the Community Aerosol and Radiation Model for Atmospheres to model cloud formation in the atmospheres of hot Jupiters. In our modeling, we consider a representative grid of planets that range in equilibrium temperature from 1300 - 2100 K with two different cases for their interior thermal structure. We also vary the amount of vertical mixing in the atmosphere and consider the impact this has on our derived cloud properties. We consider two cloud species thought to condense in this temperature range, TiO$_2$ and MgSiO$_3$. We introduce characteristic timescales of relevant processes in our model as a means to intuitively understand how the interplay between these processes influences cloud properties.

We derive fully resolved particle size distributions, total cloud masses, and vertical distributions of cloud particles for our full grid of hot Jupiters. We place these results in context by comparing the results from our modeling approach to those from other cloud models. We also calculate cloud opacities in both emission and transmission, the single scattering albedo and the asymmetry parameter of the cloud particles, and the increased accuracy obtained using a full particle size distribution as opposed to a mean particle size. These calculations allow us to determine the observational implications of our models and we compare these results to published observational inferences. Our main conclusions are summarized below.

\begin{enumerate}
\item Cloud particle size distributions are not log-normal and are instead bimodal, broad, or irregular in shape. Silicate clouds tend to be distributed broadly with an indistinct peak. Titanium clouds often have a bimodal distribution with both nucleation and growth modes. 
\item The population of titanium cloud particles is typically smaller in particle size than the population of silicate cloud particles. Titanium cloud particles are frequently abundant throughout the upper atmosphere while silicate clouds are abundant closer to their cloud base. 
\item Cloud properties depend strongly on planetary properties---in particular the temperature profile of the planet and the vertical mixing in the atmosphere. We discover a strong negative correlation between total cloud mass density and equilibrium temperature. With increased planetary equilibrium temperature, the cloud base is higher in the atmosphere and the cloud cover becomes increasingly inhomogeneous. We find that increased vertical mixing increases both the total cloud mass and the vertical extent of the clouds in the atmosphere.
\item The presence of an isothermal-like layer in planets with a low entropy interior gives rise to a deep cold trap at around 100 bar. Despite the presence of this deep cold trap, there is still significant cloud formation in the upper atmosphere. 
\item The clouds are gray across a large wavelength range in transmission and emission due to the relatively large size of the cloud particles. In both emission and transmission, the cloud opacity profile is featureless across a broad wavelength range with the exception of small features in the infrared. 
\item While the east limb has less total cloud mass than the west limb, the opaque cloud level is located higher in the atmosphere for T$_\text{eq} \le 1700$ K. The east limb therefore appears observationally to become more cloudy with increasing equilibrium temperature until the planet becomes too hot for clouds to form. Clouds form on the west limb for all planets considered in our grid.
\item Titanium and silicate clouds have strong forward scattering properties across a broad wavelength range in both transmission and emission. This indicates that a consideration of cloud scattering effects will be important when making observational inferences. 
\item A consideration of the full cloud particle size distribution leads to distinctly different cloud opacities as compared to a consideration of a mean particle size alone, often by a factor of $\sim$ 3 - 5.
\item When the full cloud particle size distribution is considered, the largest particles do not always dominate the opacity. The particle size that dominates the cloud opacity is instead dependent on the specific cloud particle size distribution. It is also possible to have a large reservoir of ``hidden" cloud mass that does not contribute to the observed cloud opacity as the cloud opacity alone is often sufficiently opaque enough to obscure the cloud base. 
\item Due to the large size of our modeled silicate clouds it is unlikely that they are responsible for the Rayleigh scattering slope in the optical---we do not see this feature in our opacity modeling. Titanium clouds are also not able to reproduce the observed Rayleigh slope.
\item In emission, at the substellar point, the cloud opacity is highly sensitive to the presence of a deep cold trap. This indicates that cloud properties may serve as useful probes of the thermal state of a planet's interior.
\end{enumerate}

This work reveals the richness and complexity involved in determining cloud properties from first principles. The results produced using bin-scheme microphysics have already changed our understanding of clouds on hot Jupiters. 

We plan to study this richness in more detail. In particular, there are three notable caveats to our modeling that we plan to address in future publications: (1) we only consider two cloud species, although other species might condense, (2) we do not consider horizontal transport of particles, and (3) we do not consider radiative feedback from clouds on the background atmospheric temperature structure. We also plan to derive full transmission spectra capable of being directly compared to observations. We are currently working to expand CARMA to 2D and eventually 3D to study the interplay between microphysics and atmospheric circulation.

\section{Acknowledgements}
We thank Jonathan Fortney, Caroline Morley, Graham Lee, Channon Visscher, Hannah Wakeford, Ruth Murray-Clay, Jamie Law-Smith, and Michael Line for their useful comments and insightful discussion. This material is based upon work supported by the National Science Foundation Graduate Research Fellowship under Grant DGE1339067.

\bibliography{refs}
\bibliographystyle{aasjournal}

\end{document}